\documentclass[preprint,showpacs,preprintnumbers,amsmath,amssymb,superscriptaddress]{revtex4}

\usepackage{graphicx,graphics}   

\usepackage{graphicx,amsfonts}
\usepackage{epsfig}

\usepackage{dcolumn}
\usepackage{bm}
\begin{document}

\title{Flavor-changing neutral currents  in the minimal 3-3-1 model revisited}

\author{A. C. B. Machado}%
\email{ana@ift.unesp.br}
\affiliation{Centro de Ci\^encias Naturais e Humanas,
Universidade Federal do ABC, Santo Andr\'e-SP, 09210-170\\Brazil
}
\author{J. C. Montero}%
\email{montero@ift.unesp.br}
\affiliation{Instituto  de F\'\i sica Te\'orica--Universidade Estadual Paulista \\
R. Dr. Bento Teobaldo Ferraz 271, Barra Funda\\ S\~ao Paulo - SP, 01140-070,
Brazil
}

\author{V. Pleitez}%
\email{vicente@ift.unesp.br}
\affiliation{Instituto  de F\'\i sica Te\'orica--Universidade Estadual Paulista \\
R. Dr. Bento Teobaldo Ferraz 271, Barra Funda\\ S\~ao Paulo - SP, 01140-070,
Brazil
}

\date{11/26/2013
}
%
\begin{abstract}
We study a few  $\Delta F=2$  and $\Delta F=1$ flavor changing neutral current processes in the minimal 3-3-1 model by considering, besides
the neutral vector bosons $Z^\prime$, the effects due to one $CP$-even and one $CP$-odd scalars. We find that there are processes in which the interference
among all the neutral bosons is constructive or destructive, and in others the interference is negligible.  We first obtain numerical values for
all the unitary matrices that rotate the left- and right-handed quarks  and give the correct
mass of all the quarks in each charge sector and the Cabibbo-Kobayashi-Maskawa (CKM) mixing matrix.
\end{abstract}


\maketitle

\section{Introduction}
\label{sec:intro}

The so called 3-3-1 models, with gauge symmetry  $SU(3)_C\otimes SU(3)_L\otimes U(1)_{X}$, are interesting extensions
of the standard model (SM). The main feature of these models is that, by choosing appropriately the representation content,
the triangle anomalies cancel out and  the number of families has to be a multiple of three, moreover
because of the asymptotic freedom this number is just three~\cite{331,331pt,mpp}. In particular, the minimal version of
this class of models (m3-3-1 for short)~\cite{331} has other interesting predictions: it  explains why $\sin^2\theta_W < 1/4$
is observed and at the same time, when $sin^2\theta_W=1/4$ it implies the existence of a Landau-like pole at energies of the
order of few TeVs~\cite{dias1}; the existence of this Landau-like pole
also stabilizes the electroweak scale avoiding the hierarchy problem~\cite{dias2}; the model allows the quantization of
electric charge independently of the nature of the massive neutrinos~\cite{pires,dong}; it also has an almost automatic
Peccei-Quinn symmetry, if the trilinear term in the scalar potential becomes a dynamical degree of freedom~\cite{axion331};
and there are also new sources of CP violation with which allow to obtain $\epsilon$ and $\epsilon^\prime/\epsilon$ even without
the CKM phase i.e., if we put $\delta=0$~\cite{cpv331}. And, probably it could explain CP violation in the $B\bar{B}$ mesons
as well. One important feature, that distinguishes the model from any other one, is the prediction of extra singly charged and
doubly charged gauge boson bileptons~\cite{dion} and also exotic charged quarks,
while the lepton sector is the same as that of the SM. Right-handed neutrinos are optional in the model. They are not needed neither
to generate light active neutrinos nor the Pontecorvo-Maki-Nakagawa-Sakata (PMNS) mixing matrix.
Those exotic charged particles may have effect on the two photon decay of the SM-like Higgs scalar~\cite{alves}.

A common feature of all 3-3-1 models is that two of the quark triplets transform differently from the third one, and this
implies flavor changing neutral currents (FCNC) at tree level, mediated by the extra neutral vector boson, $Z^\prime$,
~\cite{liu1,liu2,outras} and also by neutral scalar and pseudoscalar fields. However, in these models it is not straightforward
to put constraints on the $Z^\prime$ boson mass  from the analysis of the FCNC processes because the relevant observables
depend on unknown unitary matrices that are needed to diagonalize the quark mass matrices. Those  matrices,
here denoted by $V^{U,D}_{L,R}$, survive in some parts of the Lagrangian, in addition to their  combination
appearing in the CKM matrix, here defined as $V_{CKM}=V^U_LV^{D\dagger}_L$.

A possibility, as in \cite{promberger1}, is not to attempt to place lower bounds on the $Z^\prime$ mass, but rather set its
mass at several fixed values and  try to obtain some information about the structure of the
$V^{U,D}_{L,R}$ matrices. Moreover, usually it is considered that the dominant
contribution, by far, to FCNC is the one mediated by the $Z^\prime$, since the contributions of the (pseudo)scalars are
assumed to be negligible. Notwithstanding, we show here that this is not the most general case and there
is a range  of the parameters that allows interference between the $Z^\prime$ and, at least, one neutral scalar which
we assume as being the SM-like Higgs with a mass around 125 GeV~\cite{cmsatlas} and, at least, one pseudoscalar field.  At the LHC
energies heavy (pseudo)scalars may interfere with the $Z^\prime$ near the resonance but this will be considered elsewhere.

Our analysis implies in a new range of the parameters of 3-3-1 models that have not been
taken into account yet~\cite{godfrey,promberger1,buras12}. Another difference of our analysis from those in the literature
is that we first calculate the quark masses and all the unitary matrices appearing in the model, $V^{U,D}_{L,R}$, and which
appear, besides the usual combination $V_{CKM}$ in the charged currents with $W^\pm$, in the
Yukawa interactions. Then, we calculate the contributions of the $Z^\prime$ and the neutral (pseudo)scalar to FCNC
processes. Here we  will not consider CP violation.

We would like to emphasize that the values for the matrices $V^{U,D}_{L,R}$ should be valid at the energy scale
of the m3-3-1 model, say $\mu=\mu_{331}$. However since we do not know this energy we use instead $\mu=M_Z$. We also assume that as in the
standard model, the CKM matrix elements do not change with the energy but this has to be prove in the 3-3-1 model, and probably it is
not the case but its computation is beyond the scope of the present work. Hence, our results should be considered only as a first
illustration of the sort of analysis that have to be down in most of the extensions of the SM.

The outline of the paper is the following. In the Sec.~\ref{sec:331} we show how to obtain the $V^{U,D}_{L,R}$ matrices
by using the known quark  masses and the CKM mixing matrix.  In Sec.~\ref{sec:fcnc} we show the FCNC processes
arising at the tree level in the m3-3-1 model: those related to the $Z^\prime$ in Subsec.~\ref{subsec:nczprime}, and
those related to neutral (psudo)scalars in Subsec.~\ref{subsec:ncscalars}. Neutral processes with $\Delta F=2$ are
considered in Sec.~\ref{sec:deltaf2}: in Sec.~\ref{subsec:mk} we consider the $\Delta M_K$
and in Sec.~\ref{subsec:mb} the $\Delta M_B$ mass differences. Then, in Sec.~\ref{sec:whathiggs}, we show the conditions
under which the Yukawa interaction of the neutral scalar with mass of 125 GeV has the same coupling to the top quark as
in the SM, implying that the Higgs production mechanism is, for all practical purposes, the same in both models.
$\Delta F=1$ processes are considered in Sec.~\ref{sec:deltaf1}. The last section is devoted to our conclusions.

\section{Quark masses and mixing matrices in the minimal 3-3-1 model}
\label{sec:331}

In the 3-3-1 models of Refs.~\cite{331,331pt},  the
left-handed quark fields are chosen to form two anti-triplets
$Q^\prime_{mL}=(d_{m},\, -u_{m},j_{m})_{L}^{T}\sim({\bf
3}^{*},-1/3);\; m=1,2$; and a triplet $Q^\prime_{3L}=(u_{3},\, d_{3},\,
J)_{L}^{T} \sim({\bf 3},2/3)$. The right-handed ones are in singlets: $u_{\alpha
R}\sim({\bf 1},2/3)$, $d_{\alpha R}\sim({\bf
1},-1/3),\,\alpha=1,2,3$, $j_{mR}\sim({\bf 1},-4/3)$, and
$J_{R}\sim({\bf 1},5/3)$.
 The scalar sector, that couples to quarks,
is composed by three triplets: $\eta=(\eta^0,\,\eta^{-}_1,\,\eta^+_2)^T\sim({\bf3},0)$,
$\rho=(\rho^+,\,\rho^0,\,\rho^{++})^T\sim({\bf3},1)$ and
$\chi=(\chi^-,\,\chi^{--},\,\chi^0)^T\sim({\bf3},-1)$. Above, the numbers
between parenthesis means the transformation properties under $SU(3)_{L}$
and $U(1)_{X}$, respectively.

The model also needs a scalar sextet that gives mass to the charged leptons and neutrinos.
However, it is also possible to obtain these masses considering only the three triplets above and non-renormalizable
interactions, for details see Ref.~\cite{mpp2}. Here the effects of the sextet are in the leptonic vertex of semi-leptonic
meson decays, see Sec.~\ref{sec:deltaf1}.

With the fields above, the Yukawa interactions using the quark symmetry eigenstates are
\begin{equation}
 -\mathcal{L}_Y=\bar{Q^\prime}_{mL}[G_{m\alpha}U^\prime_{\alpha R}\rho^*+\tilde{G}_{m\alpha}
D^{\,\prime}_{\alpha R}\eta^*]+
\bar{Q^\prime}_{3L}[F_{3\alpha}U^\prime_{\alpha R}\eta+\tilde{F}_{3\alpha}D^{\,\prime}_{\alpha R}\rho]+H.c.
\label{yuka331}
\end{equation}

From Eq.(\ref{yuka331}) we obtain the following mass matrices in the basis $U^\prime_{L(R)}~=~(-u_1,-u_2,u_3)_{L(R)}$
and $D^\prime_{L(R)}=(d_1,d_2,d_3)_{L(R)}$,
\begin{eqnarray}
M^U=\left(\begin{array}{ccc}
rG_{11}&rG_{12}&rG_{13}\\
rG_{21}&rG_{22}&rG_{23}\\
F_{31}&F_{32}&F_{33}
\end{array}
\right)\vert v_\eta\vert,\;\;
 M^D=\left(\begin{array}{ccc}
r^{-1}\tilde{G}_{11}&r^{-1}\tilde{G}_{12}&r^{-1}\tilde{G}_{13}\\
r^{-1}\tilde{G}_{21}&r^{-1}\tilde{G}_{22}&r^{-1}\tilde{G}_{23}\\
\tilde{F}_{31}&\tilde{F}_{32}&\tilde{F}_{33}
\end{array}
\right)\vert v_\rho\vert.
\label{massud}
\end{eqnarray}
By choosing $\vert v_\rho\vert =54$ GeV and $\vert v_\eta\vert=240$ GeV, $r=\vert v_\rho\vert/\vert v_\eta\vert
=0.225$,
the mixing between $Z$ and $Z^\prime$ vanishes independently of the value of $\vert v_\chi\vert$ (see
the next section and Ref.~\cite{newp} for details). For simplicity we will consider all vacuum expectation values (VEVs) and
Yukawa couplings as being real numbers.

The symmetry eigenstates $U^\prime_{L,R},D^{\,\prime}_{L,R}$ and  the mass eigenstates
$U_{L,R},D_{L,R}$ (unprimed fields) are related by  $U^\prime_{L,R}=(V^U_{L,R})^\dagger U_{L,R}$ and $D^{\,\prime}_{L,R}
=(V^D_{L,R})^\dagger
D_{L,R}$, where $V^{U,D}_{L,R}$  are unitary matrices such that $V_L^UM^UV_R^{U\dagger}=\hat{M}^U$ and
$V_L^DM^DV_R^{D\dagger}=
\hat{M}^D$, where $\hat{M}^U=diag(m_u,m_c,m_t)$ and $\hat{M}^D=diag(m_d,m_s,m_b)$. The notation in these matrices
is
\begin{equation}
V^D_L=\left(\begin{array}{ccc}
(V^D_L)_{dd}& (V^D_L)_{ds} &(V^D_L)_{db}\\
(V^D_L)_{sd}& (V^D_L)_{ss} &(V^D_L)_{sb}\\
(V^D_L)_{bd}& (V^D_L)_{bs} &(V^D_L)_{bb}
\end{array}
\right),
\label{notation}
\end{equation}
for instance, and similarly for $V^D_R$ and $V^U_{L,R}$.

In order to obtain these four unitary matrices we have to solve the
matrix equations:
\begin{equation}
V^{q}_LM^qM^{q\,\dagger}V^{q\dagger}_L=
V^{q}_RM^{q\,\dagger}M^qV^{q\,\dagger}_R=(\hat{M}^q)^2,\quad q=U,D.
\label{vdvu}
\end{equation}

Solving numerically Eqs.~(\ref{vdvu}) we find the matrices $V^{U,D}_{L,R}$, which give the correct quark mass square
values and, at the same time,  the Cabibbo--Kobayashi--Maskawa quark mixing matrix (here defined as $V_{CKM}=V_L^U
V_L^{D\dagger}$). We get:
\begin{eqnarray}
&& V^U_L=\left(\begin{array}{ccc}
-0.00032 & 0.07163&  -0.99743\\
0.00433& -0.99742  & -0.07163\\
0.99999& 0.00434& -0.00001\\
\end{array}\right),\nonumber \\&&
 V^D_L\!\!=\!\!\left(\begin{array}{ccc}
0.00273 \to0.00562& (0.03\to0.03682)& -(0.99952\to0.99953)\\
-(0.19700\to0.22293)& -(0.97436\to0.97993) & -0.03052\\
0.97483\to0.98039& -(0.19708\to0.22291)& -(0.00415\to 0.00418)\\
\end{array}\right)
\label{vudl331},
\end{eqnarray}
and the CKM matrix
\begin{equation}
\vert V_{CKM}\vert =\left(\begin{array}{ccc}
0.97385\to0.97952 & 0.20134\to0.22714& 0.00021\to0.00399\\
0.20116\to0.22679&0.97307\to0.97869 &0.04116\to0.04118\\
0.00849\to0.01324 & 0.03919\to0.04028 & 0.99914\to0.99915\\
\end{array}\right),
\label{ckm2}
\end{equation}
which is in agreement with the data~\cite{pdg}.
In the same way we obtain the $V^{U,D}_R$ matrices:
\begin{eqnarray}
&& V^U_R=\left(\begin{array}{ccc}
-0.45440 & 0.82278 &  -0.34139\\
0.13857& -0.31329 &-0.93949 \\
0.87996& 0.47421& -0.02834\\
\end{array}\right),\nonumber  \\&&
 V^D_R\!\!=\!\!\left(\begin{array}{ccc}
-(0.000178\to0.000185)& (0.005968 \to0.005984 )& -0.999982\\
-(0.32512\to0.32559)& -(0.94549\to0.94566) & -(0.00558\to0.00560)\\
0.94551\to0.94567& -(0.32511\to0.32558)& -(0.00211-0.00212 )\\
\end{array}\right).
\label{vudr331}
\end{eqnarray}

The values for the coupling constants in Eq.~(\ref{massud}), which give the numerical values for the matrix entries
in Eqs.~(\ref{vudl331})-(\ref{vudr331}), are:  $G_{11}=1.08,G_{12}=2.97,G_{13}=0.09,G_{21}=0.0681,
G_{22}=0.2169,G_{23}=0.1\times10^{-2}$,
$F_{31}=9\times10^{-6},F_{32}=6\times10^{-6},F_{33}=1.2\times10^{-5}$, $\tilde{G}_{11}=0.0119,
\tilde{G}_{12}=6\times10^{-5},\tilde{G}_{13}=2.3\times10^{-5},\tilde{G}_{21}=(3.62 - 6.62)\times10^{-4},\tilde{G}_{22}=
2.13\times10^{-4},\tilde{G}_{23}=7\times10^{-5}$, $\tilde{F}_{31}=2.2\times10^{-4},
\tilde{F}_{32}=1.95\times10^{-4},\tilde{F}_{33}=1.312\times10^{-4}$. With the values above we obtain
from Eq.~(\ref{vdvu}) the masses at the $Z$ pole (in GeV): $m_u=0.00175,m_c=0.6194,m_t=171.163$, and
$m_d=(33.6 - 39.3)\times10^{-4},m_s= (0.0544 - 0.0547),m_b=(2.8537-2.8574)$ which are in agreement
with the values given in Ref.~\cite{massas}. For the sake of simplicity, we are only allowing the $d$-type quark
masses to vary within the 3$\sigma$ experimental error range. These results are valid for the models in
Refs.˜\cite{331,331pt} but other 3-3-1 models can be similarly studied.

\section{Neutral current interactions}
\label{sec:fcnc}

It is usually considered that 3-3-1 models reduce to the SM only at high energies. If $v_\chi$
is the VEV that breaks the 3-3-1 symmetry down to the 3-2-1 one, then $v_\chi\gg~v_{SM}~=~(1/\sqrt{2}G_F)^{1/2}
\approx246$ GeV. In this limit the lightest neutral vector boson, $Z_1$, whose mass is $M_{Z_1}$, has for all practical
purposes the same couplings with fermions of the SM $Z$, since in this case the mixing among $Z$ and $Z^\prime$ is
less than $10^{-3}$~\cite{outras}. This mixing is small due to the existence of an approximate $SU(2)_{L+R}$ custodial
global symmetry, see Ref.~\cite{newp}.

However, there is another solution which also reproduces the SM model couplings for the lightest neutral vector boson,
$Z_1$, without imposing  $v_\chi\gg v_{SM}$
at the very star. This is a non trivial solution that
implies that $Z$ and $Z^\prime$ do not mix, at the tree level, independently of the value of $v_\chi$ as it was shown
in Ref.~\cite{newp}. There, it is defined the $\rho_1$--parameter as $\rho_1=c_W^2M_{Z_1}^2/M_W^2$, where
$M_{Z_1}$ has a  complicate dependence on all the VEVs and $\sin^2\theta_W$. In general $\rho_1\leq1$ since
$M_{Z_1}\leq M_Z$. In the SM context it is defined $\rho_0=c^2_WM^2_Z/M^2_W$. We define the SM limit of the
3-3-1 model, at the tree level, imposing the condition $\rho_1=\rho_0=1$.
We find that this condition is satisfied in two cases: first, the usual one when $v_\chi \to \infty$. A second non trivial
solution for satisfying this condition can be found  by solving for $v_\rho=\sqrt{2}\langle \rho^0\rangle$, given the
solution $\bar{v}_\rho^2=[(1-4s_W^2)/2c_W^2]\bar{v}_{SM}^2$, where $\bar{v}_\rho=v_\rho/v_\chi$, and
$\bar{v}_{SM}=v_{SM}/v_\chi$. As $v_\rho$ and $v_\eta$
($v_\eta=\sqrt{2}\langle\eta^0\rangle)$ are constrained by $v_{SM}$ as $v_\rho^2+v_\eta^2=v_{SM}^2$, in order
to give the correct mass to $M_W$, we find $\bar{v}_\eta^2=[(1+2s_W^2)/2c_W^2]\bar{v}_{SM}^2$, where
$\bar{v}_\eta=v_\eta/v_\chi$. It implies that the VEVs of the triplets
$\eta$ and $\rho$ must have the values considered in the previous section i.e.,
$\vert v_\rho\vert =54$ GeV and $\vert v_\eta\vert=240$ GeV, while leaving $v_\chi$ completely
free, and it may be even of the order of the electroweak scale, unless there are constraints coming from specific experimental
data. This justify the values for these VEVs used in Eqs.~(\ref{massud}) and (\ref{vdvu}).

The nontrivial solution above is in fact the SM limit of the 3-3-1 model:
When the expressions of $\bar{v}_\rho$ and $\bar{v}_\eta$ are used in the full expression of $M_{Z_1}$, we obtain
that $M_{Z_1}=M_Z$. This also happens with the couplings of $Z_1$ to the known fermions, denoted generically by  $i$,
say $g^{Z_1,\;i}_{V,A}$, which in this model are also complicated functions of all VEVs and $\sin^2\theta_W$. It is found
that they  are exactly the same as the respective couplings of the  SM's $Z$, $g^{Z_1,\;i}_{V,A}\to g^{Z\,SM,\,i}_ {V,A}$.
Moreover, this SM limit is obtained regardless the $v_\chi$ scale, since it factorizes in both sides of the relations defining
$\bar{v}_\rho$. In any case the $Z^\prime$, with a mass that depends mainly on $v_\chi$9 may be
lighter than we thought before if $v_ \chi \stackrel{>}{\sim} v_{SM}$. From this SM limit it results that
$M_{Z_1}\equiv M_Z$, \, $Z_1\equiv Z$, and $Z_2\equiv Z^\prime $, and there is no mixing at all between $Z$
and $Z^\prime$ at the tree level. See Refs.~\cite{newp} for details.

A light $Z^\prime$ is then a theoretical possibility. However, the phenomenology of FCNC may impose strong
lower bounds on $M_{Z^\prime}$.  Here we will consider  FCNC processes induced by both, $Z^\prime$
and neutral scalars and pseudoscalars. In some of these processes there is non-negligible interference among
all neutral particle contributions and, depending on a given range of the parameters, the interference may be
constructive or destructive. This sort of interference happens when at least one
neutral scalars, with mass of the order of the 125 GeV and/or a pseudoscalar with a mass larger  than the scalar
one are considered. The (pseudo)scalars
have to be included since their interactions with quarks are not proportional to the quark masses. In the next
subsections we show explicitly the quark neutral current interactions  which induce FCNC, for both the
$Z^\prime$ and scalar fields.

\subsection{Neutral currents mediated by the $Z^\prime$}
\label{subsec:nczprime}

As it is well known~\cite{liu1,liu2,outras}, the neutral vector boson $Z^\prime$ induces FCNC at the tree level.
In fact, its interactions to quarks are given by the Lagrangian
\begin{eqnarray}
\mathcal{L}_{Z^\prime}&=&-\frac{g}{2\cos\theta_W}\sum_{q=U,D}[\bar{q}_L\gamma^\mu K^q_L q_L+
\bar{q}_R\gamma^\mu K^q_R q_R]Z^\prime_\mu,
\label{zprime}
\end{eqnarray}
where we have defined
\begin{equation}
K^q_L=V^{q}_LY^q_LV^{q\dagger}_L,\;\; K^q_R =V^{q}_RY^q_RV^{q\dagger}_R,\;\;q=U,D;
\label{kas}
\end{equation}
with $V^{U,D}_{L,R}$  given in Eqs.~(\ref{vudl331}) and (\ref{vudr331}) and
\begin{equation}
Y^U_L=Y^D_L=-\frac{1}{2\sqrt{3}h(x)}\textrm{diag}\,[ -2(1-2x),-2(1-2x),1]
\label{yudl}
\end{equation}

and
\begin{equation}
Y^U_R=-\frac{4x}{\sqrt{3}h(x)}\,\textbf{1}_{3\times3} ,\quad
Y^D_R=\frac{2x}{\sqrt{3}h(x)}\textbf{1}_{3\times3}
\label{yudr}
\end{equation}
and $h(x)\equiv(1-4x)^{1/2},\;x=\sin^2\theta_W$. See Ref.~\cite{dpp}.

Using the matrices in Eqs.~(\ref{vudl331}), (\ref{yudl}) and (\ref{yudr}) we obtain for the $K^q_L$ defined
in Eqs.~(\ref{kas})
\begin{eqnarray}
&&
K^U_L\approx\left(
\begin{array}{ccc}
-1.04793&-0.08905&-0.00004  \\
-0.08905&1.12718&-10^{-6}\\
-0.00004&-10^{-6}&1.13088
\end{array}
\right),\nonumber \\&&
K^D_L\approx\left(
\begin{array}{ccc}
-1.05154 &-0.00140 &-0.00826\\
-0.00140  &1.13082 &-5\cdot10^{-6}  \\
-0.00826&-5\cdot10^{-6} &1.13078
\end{array}
\right).
\label{numeros}
\end{eqnarray}
Since $Y^{U,D}_R$ are proportional to the identity matrix, there are no FCNCs in the right-handed currents
coupled to the $Z^\prime$, and using the matrices in Eqs.~(\ref{vudr331}) we obtain $K^U_R\approx-1.94465
\,\textbf{1}_{3\times3}$ and $K^D_R\approx0.97232\,\textbf{1}_{3\times3}$.

\subsection{Neutral currents mediated by scalars and pseudoscalars}
\label{subsec:ncscalars}

As we said before, there are also FCNCs at the tree level in the scalar sector. From Eq.~(\ref{yuka331}) we obtain
the following neutral scalar-quark interactions
\begin{equation}
-\mathcal{L}_{qqh}=\sum_{q=U,D}\overline{q}_L\mathcal{K}^q q_R
+\textrm{mass terms}+H.c. ,
\label{yukan}
\end{equation}
where we have defined $\mathcal{K}^U= V^U_L\mathcal{Z}^UV^{U\dagger}_R$ and $\mathcal{K}^D= V^D_L\mathcal{Z}^D
V^{D\dagger}_R$ and we have arranged,
for simplicity, the interactions in matrix form (in the quark mass eigenstates  basis):
\begin{eqnarray}
\mathcal{Z}^U=\left(\begin{array}{ccc}
G_{11}\rho^0 &G_{12}\rho^0&G_{13}\rho^0\\
G_{21}\rho^0&G_{22}\rho^0&G_{23}\rho^0\\
F_{31}\eta^0&F_{32}\eta^0&F_{33}\eta^0
\end{array}
\right),\;\;
\mathcal{Z}^D=\left(\begin{array}{ccc}
\tilde{G}_{11}\eta^0&\tilde{G}_{12}\eta^0&\tilde{G}_{13}\eta^0\\
\tilde{G}_{21}\eta^0&\tilde{G}_{22}\eta^0&\tilde{G}_{23}\eta^0\\
\tilde{F}_{31}\rho^0&\tilde{F}_{32}\rho^0&\tilde{F}_{33}\rho^0
\end{array}
\right),
\label{yukanma}
\end{eqnarray}
where $\eta^0$ and $\rho^0$ are still symmetry eigenstates. These neutral symmetry
eigenstates  may be written as $\sqrt{2}x^0=\textrm{Re}
\,x^0+i\textrm{Im}\,x^0$, with $x^0=\eta^0,\rho^0$, and their relations to mass eigenstates are defined as
$\textrm{Re}\,\eta^0=\sum_iU_{\eta i}h^0_i,\;\textrm{Re}\,\rho^0=\sum_iU_{\rho i}h^0_i$. The real scalars
$h^0_i$ are mass eigenstates
with mass $m_i$, and similarly for the imaginary part pseudoscalar fields, $\textrm{Im}\,\eta^0=\sum_i
V_{\eta i}A^0_i$ and $\textrm{Im}\,\rho^0=\sum_i V_{\rho i}A^0_i$,
with $A^0_i$ including two Goldstone bosons and at least one $CP$-odd mass eigenstate. The physical
$CP$-odd pseudoscalars have a mass denoted $m_{A_i}$.
Notice that, besides  the matrices $U_{\eta1}$ and $U_{\rho1}$, the matrices $V^{U,D}_{L,R}$ survive in the
interactions given in Eqs.~(\ref{yukan}) and (\ref{yukanma}).

Using in Eq.~(\ref{yukanma}) the values of $G,F,\tilde{G},\tilde{F}$ written below  Eq.~(\ref{vudr331}),  and
the matrices $V^D_L$ and $V^D_R$, given  in Eqs.~(\ref{vudl331}) and (\ref{vudr331}), respectively,  the matrix
$\mathcal{K}^D$ in Eq.~(\ref{yukan}) is given by
\begin{eqnarray}
&&\mathcal{K}^D\approx\left(
\begin{array}{ccc}
10^{-4}\rho^0 -10^{-6}\eta^0 & 10^{-4}\rho^0-10^{-5}\eta^0 & -10^{-4}\rho^0+10^{-5}\eta^0\\
10^{-6}\rho^0+10^{-4}\eta^0&10^{-5}\rho^0+10^{-3}\eta^0 &-10^{-6}\rho^0+10^{-2}\eta^0\\
10^{-6}\rho^0-10^{-5}\eta^0&10^{-6}\rho^0-10^{-3}\eta^0&-10^{-6}\rho^0+0.011\eta^0
\end{array}
\right),
\label{kapasd}
\end{eqnarray}
where we have shown only the order of magnitude of each entry. As many multi-Higgs extensions of the SM,
in the m3-3-1 model there are other scalars that mix with the SM-like Higgs boson. These scenarios may be tested
experimentally if the couplings of the 125 GeV Higgs are measured~\cite{gupta,holdom}.

In the present model the interaction vertex $h^0_1VV$, $V=W,Z$ include all the scalar components of the neutral scalars
and pseudoscalars which couple to the known quarks and leptons i.e., this vertex is
proportional to $y_V(\sum_i U_{i1})$, $i=\eta,\rho,\sigma_2,\sigma_1$ where $\sigma_1$ and $\sigma_2$ denote the neutral
components in the scalar sextet, and $y_V$ is the respective vertex in the SM. If
$\sum_iU_{i1}\stackrel{<}{\sim}1$ we can have agreement with the SM strength.
On the other hand, the interactions with fermions have additional reducing factors given by the numbers $\mathcal{K}^{U,D}_{q_1q_2}$ in
Eq.~(\ref{kapasd}) and (\ref{kapasu}) below.
In this case the strength of the couplings are given, for instance, by the matrix
elements of $\mathcal{K}^D$, in Eq.~(\ref{kapasd}),
with $\eta^0\to U_{\eta1}h^0_1$ and $\rho^0\to U_{\rho1}h^0_1$.
We will denote $\mathcal{K}^{Dx}_{q_1q_2}=\mathcal{\bar{K}}^D_{q_1q_2}U_{x1},\;x=\eta,\rho$, where
$\mathcal{\bar{K}}^D_{q_1q_2}$ denotes the number
in the respective entries in $\mathcal{K}^D$, and
similarly with the pseudoscalar $A^0_1$, although the latter one has no counterpart in the SM.
Notice that in the present model, the diagonal elements are $(\mathcal{K}^{D\rho})_{dd}\approx 1
0^{-4}U_{\rho1}$, $(\mathcal{K}^{D\eta})_{ss}\approx
10^{-3}U_{\eta1}$,
and $(\mathcal{K}^{D\eta})_{bb}\approx 1.1\times 10^{-2}U_{\eta1}$.

In the SM, the neutral scalar has only diagonal interaction to a fermion $f$:
$y_f=\frac{m_f\sqrt{2}}{v}$, hence we have the following Yukawa couplings $y_d=2.8\times10^{-5}$,
$y_s=5.5\times10^{-4}$ and $y_b=2.7\times10^{-2}$.
Hence, since $\vert U_{x 1}\vert\leq 1$, for
$x =\eta,\rho$, we see that the quarks $d$ and $s$ can
have the same numerical
Yukawa couplings as in the SM, but this is not the case for the $b$-quark since $(\mathcal{K}^{D\eta})_{bb}
\stackrel{<}{\sim}y_b$ even if $\vert U_{\eta1}\vert=1$.
We recall that this happens at the energy scale $\mu=M_Z$ and, it is not obvious that in this model these couplings do not change
enough between this energy and 125 GeV.

Notwithstanding, at present we have to compare this value not with the SM one but with the measured
value  and the respective errors. Denoting $y_{bbh}=y_b(1+\Delta_b)$ experimental data still allow
$1.04\times10^{-2}<y_{hhb}<4.6 \times10^{-2}$~\cite{plehn} and we see that
$y_{bbh}$ is still compatible with the value of $(\mathcal{K}^{D\eta})_{bb}$ above.
Recently,  the first indication of the $H\to bb$ decay at the LHC has been obtained
by the CMS collaboration. It has an excess of $2.1\sigma$ relative to that of the SM Higgs boson~\cite{cmsbbar}.
On the other hand, fermiophobic scalars have been excluded in the mass ranges 110.0 -- 118.0 GeV and 119.5 -- 121.0
GeV~\cite{atlas} but, in fact, the important coupling of a Higgs with mass of the order of 125 GeV is that with
the $t$-quark, see also Sec.~\ref{sec:whathiggs}. The present model corresponds under
the $SU(2)_L\otimes U(1)_Y$ subgroup, to a model with three-Higgs doublets $Y=+1$, a neutral scalar singlet $Y=0$, and a non-Hermitian
triplet $Y=2$ which couples to leptons. The latter Higgs with $Y=2$ belongs to a $SU(3)$ sextet. We will consider only the two triplets
which couple to quarks and assume that one of the scalar mass eigenstates has a mass consistent with the recent results from LHC,
$m_1=125$ GeV~\cite{cmsatlas}  and a pseudoscalar with mass $m_A$. We use some FCNC processes to get constraints on $M_{Z^\prime}$, $U_{x1}$,
$V_{x1}$ ($x=\eta,\rho$) and $m_A$.


\section{$\Delta F=2$ processes 
}
\label{sec:deltaf2}

In 3-3-1 models, $\Delta F=2$  transitions ($F=S,B,C$) at tree and loop level arise. In this section we will consider only
the strange and beauty cases. The $D_0-\bar{D}^0$ will be consider in Sec.~\ref{sec:whathiggs}.
The main contributions to these processes are those at the tree level and they are mediated by $Z^\prime$ and
neutral (pseudo-)scalars.

\subsection{$\Delta M_K$}
\label{subsec:mk}

In the SM context, the $\Delta M_K$ mass difference in the neutral kaon system is given by
$\Delta M^{SM}_K=\zeta^{SM}_{sd}\langle \bar{K}^0\vert (\bar{s}d)^2_{V-A} \vert K^0\rangle$ where,
using only the $c$-quark contribution, we have
\begin{equation}
\zeta^{SM}_{sd}=\frac{G^2_Fm^2_c}{16\pi^2}\,[(V_{CKM})^*_{cd}(V_{CKM})_{cs}]^2\approx 10^{-14}\,\textrm{GeV}^{-2},
\label{deltaksm}
\end{equation}
and we have neglected QCD corrections and in the vacuum insertion approximation
we have $\langle \bar{K}^0\vert (\bar{s}d)^2_{V-A} \vert K^0\rangle=\frac{1}{3}M_Kf^2_K$~\cite{branco}.

Let us consider first the contributions of the extra neutral vector boson.
From Eq.~(\ref{zprime}), the effective $Z^\prime$ interaction Hamiltonian inducing the $K^0 \to \bar{K}^0$ transition,
at the tree level, is given by
\begin{eqnarray}
\mathcal{H}^{\Delta S=2}_{eff}\vert_{Z^\prime}=\frac{g^2}{ 4 c^2_W M^2_{Z^\prime}} [\bar{s}_L (K^D_L)_{sd}
\gamma^\mu d_L]^2,
\label{heff}
\end{eqnarray}
and we obtain the following extra contribution to $\Delta M_K$
\begin{eqnarray}
\Delta M_K\vert_{Z^\prime}=2\textrm{Re}\,\langle \bar{K}^0\vert \mathcal{H}^{\Delta S=2}_{eff}
\vert_{Z^\prime}\vert K^0\rangle= \textrm{Re}\,\zeta^{Z^\prime}_{sd}\,\langle \bar{K}^0\vert (\bar{s}d)^2_{V-A}
\vert K^0\rangle
\label{deltak}
\end{eqnarray}
where
\begin{equation}
\textrm{Re}\,\zeta^{Z^\prime}_{sd}=\textrm{Re}\,\frac{G_F}{2 \sqrt{2} c^2_W}\,\frac{M^2_W}{M^2_{Z^\prime}}
\;[(K^D_L)_{ds}]^2
\,= \,
\frac{M^2_W}{M^2_{Z^\prime}}\,10^{-11}\,\textrm{GeV}^{-2},
\label{deltak331}
\end{equation}
since, from Eq.~(\ref{numeros}), we have $(K^d_L)_{sd}=-1.4 \times10^{-3}$. If this were the only contribution
to $\Delta M_K$,
and imposing $\zeta^{Z^\prime}_{sd}<\zeta^{SM}_{sd}$, we must have that $M_{Z^\prime}>2.5$ TeV.

Next, let us consider the scalar contributions to $\Delta M_K$.
From Eq.~(\ref{yukan}), the scalar interactions between the  $d$ and $s$ quarks mediated by $h^0_i$  are given by
\begin{eqnarray}
-\mathcal{L}_{dsh}&=&
\frac{1}{\sqrt2}\sum_i[(I^{i}_K)_{ds} \bar{s_L}d_Rh^0_i+ (I^{i*}_K)_{sd}\bar{d}_Ls_Rh^0_i+H.c.]
\nonumber\\&=&\frac{1}{2 \sqrt2} \sum_i[(I^{i+}_K)_{ds}(\bar{d}s)+(I^{i-}_K)_{ds}(\bar{d}\gamma_5s)]h^0_i+ H.c.,
\label{newint}
\end{eqnarray}
where, $(I^i_K)_{q_1q_2} = (\mathcal{K}^D)_{q_1q_2} U_{x i}$ with $x=\eta,\rho$ and $q_1,q_2=d,s$ for the real
scalars and quarks, respectively;
$i$ run over the neutral scalar mass eigenstates and the matrix $\mathcal{K}^D$ is defined in (\ref{kapasd}) and in
the second line of (\ref{newint})
we have defined  $(I^{i\pm}_K)_{ds} =(I^{i}_K)_{ds} \pm (I^{i*}_K)_{sd}$. For $CP$-odd fields the
Lagrangian is similar to that in (\ref{newint}), but with $h^0_i\to A^0_i$ and $(I^i_K)_{q_1q_2} \to(I^i_K)^A_{q_1q_2}
= (\mathcal{K}^D)_{q_1q_2} V_{x i}$. For the definition of $U_{x i}$ and $V_{x i}$ see the discussion below
Eq.~(\ref{yukanma}). Then, using the numbers in Eq.~(\ref{kapasd}), we have
\begin{equation}
(I^{i}_K)_{ds}\approx  10^{-4} U_{\rho i}-10^{-5}U_{\eta i },\quad
(I^i_K)_{sd}\approx  10^{-6}U_{\rho i}+10^{-4} U_{\eta i }.
\label{ies}
\end{equation}
For the pseudoscalar contributions $(I^i_K)^A$ are the same as in Eq.~(\ref{ies}) but with $U_{\eta i}\to V_{\eta i}$ and
$U_{\rho i}\to V_{\rho i}$.

The effective Hamiltonian induced by Eq.~(\ref{newint}), and the respective contribution of the pseudoscalar $A^0_1$
to the $K^0 \leftrightarrow \bar{K}^0$ transition is:
\begin{eqnarray}
\mathcal{H}^{\Delta S=2}_{eff}\vert_{h+A}&=&\sum_i\frac{1}{8 m^2_i}[(I^{i+}_K)^2_{ds}(\bar{s}d)^2+(I^{i-}_K)^2_{ds}(
\bar{s}\gamma_5d)^2]\nonumber \\&-&
\sum_i\frac{1}{8 m^2_A}[[(I^{i+}_K)^A_{ds}]^2(\bar{s}d)^2+[(I^{i-}_K)^A_{ds}]^2(\bar{s}\gamma_5d)^2.
\label{heff2}
\end{eqnarray}
Defining as usual
\begin{equation}
\Delta M_K\vert_{h,A}=2
\langle \bar{K}^0\vert
\mathcal{H}^{\Delta S=2}_{eff}\vert_{h,A} \vert K^0\rangle=\textrm{Re}\zeta^{h,A}_{sd}\langle \bar{K}^0
\vert (\bar{s}d)^2_{V-A} \vert K^0\rangle,
\label{effective2}
\end{equation}
and  using the matrix elements~\cite{branco}:
\begin{eqnarray}
&&\langle \bar{K}^0\vert (\bar{s}d)(\bar{s}d)\vert K^0\rangle=-\frac{1}{4}\left[1-\frac{M^2_K}{(m_s+m_d)^2}
\right] \langle
\bar{K}^0\vert (\bar{s}d)^2_{V-A} \vert K^0\rangle,\nonumber \\&&
\langle \bar{K}^0\vert (\bar{s}\gamma_5d)(\bar{s}\gamma_5d)\vert K^0\rangle=\frac{1}{4}\left[1-1
1\frac{M^2_K}{(m_s+m_d)^2}
\right]\langle \bar{K}^0\vert (\bar{s}d)^2_{V-A} \vert K^0\rangle,
\label{v1v2}
\end{eqnarray}
we find
\begin{eqnarray}
\textrm{Re}\zeta^h_{ds}\!\!=\!\!\textrm{Re}\sum_i\frac{1}{32 m^2_i}\left(-(I^{i+}_K)^2_{ds}\left[1\!\!-
\!\!\frac{M^2_K}{(m_s+m_d)^2} \right]\!\!
+\!\!(I^{i-}_K)^2_{ds}\left[1\!\!-\!\!\frac{11M^2_K}{(m_s+m_d)^2} \right] \right)\textrm{GeV}^{-2},
\label{zetas2}
\end{eqnarray}
and
\begin{equation}
(I^{i\pm}_K)^2_{ds} \approx [(10U^*_{\rho i}-2U^*_{\eta i})U^*_{\rho i} \pm (10 U^*_{\rho i}-
U^*_{\eta i})U_{\eta i}+10(U_{\eta i})^2 ]\times 10^{-9}.
\label{i+-}
\end{equation}

Then, Eq.~(\ref{zetas2}) becomes
\begin{eqnarray}
&&\textrm{Re}\zeta^h_{ds}=\textrm{Re}\sum_i\frac{1}{32 m^2_i} [24
[(10U^*_{\rho i}-2U^*_{\eta i})U^*_{\rho i} + (10 U^*_{\rho i}- U^*_{\eta i})U_{\eta i}+10(U_{\eta i})^2]
\nonumber \\&&
-272[
(10U^*_{\rho i}-2U^*_{\eta i})U^*_{\rho i} - (10 U^*_{\rho i}- U^*_{\eta i})U_{\eta i}+10(U_{\eta i})^2
]
 \times\,10^{-9}\,\textrm{GeV}^{-2}.
\label{zetas3}
\end{eqnarray}

We have similar expressions for the pseudoscalar contributions by making, in Eq.~(\ref{zetas3}),
$I^{i\pm}_K\to (I^{i\pm})^A_K$, with
$U_{\eta1}\to V_{\eta1}$, $U_{\rho1}\to V_{\rho1}$ and $m_i\to m_{Ai}$.  Thus, the $\Delta M_K$ in the present
model includes $Z^\prime$
and neutral scalar and pseudoscalar contributions,
\begin{equation}
\Delta M_K\vert_{331}\approx \Delta M^{SM}_K+
\Delta M_K\vert^{Z^\prime}+\Delta M_K\vert^h+\Delta M_K\vert^A\equiv\zeta_{331}\langle \bar{K}^0
\vert (\bar{s}d)^2_{V-A} \vert K^0\rangle,
\label{full}
\end{equation}
with $\zeta_{331}=\zeta^{SM}_{ds}+\zeta^{Z^\prime}_{ds}+\zeta^h_{ds}+\zeta^A_{ds}$, and we impose that
$\zeta^{Z^\prime}_{ds}+\zeta^h_{ds}+\zeta^A_{ds}<\zeta^{SM}_{ds}$, hence
\begin{equation}
\textrm{Re}\,(\zeta^{Z^\prime}_{ds}+\zeta^h_{ds}+\zeta^A_{ds}) <10^{-14}\,\textrm{GeV}^{-2}.
\label{effective3}
\end{equation}

Using Eqs.~(\ref{deltak331}) and (\ref{zetas3}) in Eq.~(\ref{effective3}), and assuming that only one of the
SM-like neutral Higgs (pseudo)scalar
contribute, say $h^0_1$ and $A^0_1$ (the others are considered too heavy and their contributions can be
neglected), Eq.~(\ref{effective3}) becomes
\begin{eqnarray}
&& 10^{-2}\,
\frac{M^2_W}{M^2_{Z^\prime}}\,+\textrm{Re}\frac{1}{32 m^2_1} \left\{24([
(10U^*_{\rho 1}-2U^*_{\eta 1})U^*_{\rho 1} + (10 U^*_{\rho 1}- U^*_{\eta 1})U_{\eta 1}+10(U_{\eta 1})^2
\right.\nonumber \\&&\left.
-272[
(10U^*_{\rho 1}-2U^*_{\eta 1})U^*_{\rho 1} - (10 U^*_{\rho 1}- U^*_{\eta 1})U_{\eta 1}+10(U_{\eta 1})^2]
 \right\}-\mathcal{A} <10^{-5}\,\textrm{GeV}^{-2},
\label{ok}
\end{eqnarray}
where $\mathcal{A}$ is the amplitude induced by the pseudoscalar $A^0_1$, which is similar to the scalar
one in Eq.~(\ref{ok}) but with $m_1\to m_A$,
$U_{\eta1}\to V_{\eta1}$ and $U_{\rho1}\to V_{\rho1}$. Once we are considering a SM-like neutral scalar
its mass $m_1$ is fixed in 125 GeV and $m_A$
is free. Hence, in Eq.~(\ref{ok}) the only free parameters are the masses of $Z^\prime$ and $A^0_1$, and
the matrix elements $U_{ \eta1},U_{\rho1}$ and
$V_{ \eta1},V_{\rho1}$.

Fist, we will not consider in (\ref{ok}) the pseudoscalar $A^0_1$, assuming that $U_{ \eta1}$ and $U_{\rho1}$
are real and within the interval $[-1,1]$. Next,
we will keep $U_{\rho1}$ fixed and varying $U_{\eta1}$, we obtain the corresponding $Z^\prime$ mass which
satisfies Eq.~(\ref{ok}) that runs from GeVs
to few TeVs. See the curves in Figs.~\ref{fig1}-\ref{fig3} and discussion in Sec.~\ref{sec:results}.

In the next subsection we will consider FCNC processes as in the previous one but now involving the $b$ quark.

\subsection{$\Delta M_B$}
\label{subsec:mb}

We can also consider the $B^0_d-\bar{B}^0_d$ mass difference, $\Delta M^{SM}_B=\zeta^{SM}_{bd}\langle
\bar{B}^0\vert (\bar{s}d)^2_{V-A}
\vert B^0\rangle$
where $\langle \bar{B}^0\vert (\bar{b}d)^2_{V-A} \vert B^0\rangle=M_Bf^2_B/3$~\cite{buras},
and, as before, we factorized the model independent factors
\begin{equation}
\zeta^{SM}_{bd}=\frac{G^2_FM^2_W}{12\pi^2}\,S_0(x_t)[(V_{CKM})^*_{td}(V_{CKM})_{tb}]^2\approx  1.0329
\times 10^{-12}\,\textrm{GeV}^{-2},
\label{deltabsm}
\end{equation}
where $x_t=m^2_t/M^2_W$ and we have used $S_0(x)\approx 0.784x_t^{0.76}$~\cite{schneider}.

From Eqs.~(\ref{zprime})-(\ref{yudr})
the effective Hamiltonian contributing to the $B^0_d\leftrightarrow \bar{B}^0_d$ transition is given by
\begin{eqnarray}
\mathcal{H}^{\Delta B=2}_{eff}\vert_{Z^\prime}=\frac{g^2}{4 c^2_W M^2_{Z^\prime}} [\bar{b}_L (K^D_L)_{bd}
\gamma^\mu d_L]^2,
\label{heff3}
\end{eqnarray}
and we obtain the following extra contributions to $\Delta M_{B_d}$, using here and below, appropriate matrix
elements as in Eq.~(\ref{v1v2}) for the kaon system,
\begin{eqnarray}
\Delta M_{B_d}\vert_{Z^\prime}=2\textrm{Re}\,\langle \bar{B}^0\vert \mathcal{H}^{\Delta B=2}_{eff}
\vert_{Z^\prime}\vert B^0\rangle= \textrm{Re}\,\zeta^{Z^\prime}_{bd}\,\langle \bar{B}^0\vert (\bar{b}d)^2_{V-A}
\vert B^0\rangle
\label{deltab}
\end{eqnarray}
where, and we have not considered the QCD corrections and the bag parameter $B_B=1$. We obtain
\begin{equation}
\textrm{Re}\,\zeta^{Z^\prime}_{bd}=\textrm{Re}\,\frac{G_F}{ 2 \sqrt{2} c^2_W}\,\frac{M^2_W}{M^2_{Z^\prime}}
\;[(K^D_L)_{bd}]^2
\,=10^{-9}\,
\frac{M^2_W}{M^2_{Z^\prime}}\,\textrm{GeV}^{-2},
\label{deltab331}
\end{equation}
where we have used (\ref{numeros}), i.e., $(K^D_L)_{bd}=-8.3 \times 10^{-3}$.

Similarly we have the scalar contributions in the $B^0_q-\bar{B}^0_q$ system ($q=d,s$). From Eqs.~(\ref{yukan})
and (\ref{kapasd})
the scalar interactions between the  $b$, $d$ quarks mediated by the scalars $h^0_i$ are given by
\begin{eqnarray}
-\mathcal{L}_{bqh}&=&
\frac{1}{\sqrt2}\sum_i[(I^i_{B_d})_{bq} \bar{b}_L d_R+ (I^{\prime i}_{B_d})_{bq}\bar{d}_Lb_R]h^0_i+H.c.
\nonumber\\&=&\frac{1}{2\sqrt2} \sum_i[(I^{i+}_{B_d})_{bq}(\bar{b}d)+(I^{i-}_{B_d})_{bq}
(\bar{b}\gamma_5d)]h^0_i+ H.c.,
\label{newint2}
\end{eqnarray}
where $(I^i_{B_d})_{q_1q_2}=(\mathcal{K}^D)_{q_1q_2}U_{\alpha i};\;\alpha=\eta,\rho;\;q_i,q_2=b,d$.
The respective entries of the matrix $\mathcal{K}^D$ can be obtained from Eq.~(\ref{kapasd}).
We have defined $(I^{i\pm}_B)_{bq}=
(I^{\prime i}_{B_d})_{bq} \pm (I^{i*}_{B_d})_{qb}$.  For the case when $q=d$  we obtain
\begin{eqnarray}
(I^i_{B_d})_{bd}\approx 10^{-6}U_{\rho i}- 10^{-5} U_{\eta i } ,
\,\,
(I^{i}_{B_d})_{db}\approx  -10^{-4} U_{\rho i}+ 10^{-5}U_{\eta i } .
\label{ies2}
\end{eqnarray}
The contributions of the pseudoscalar fields are similar to those of the scalar $h^0_i$ but making
$h^0_1\to A^0_i$ in (\ref{newint2}),  and
$U_ {\eta i}\to V_{\eta i}$ and $U_{\rho i}\to V_{\rho i}$ in Eq.~(\ref{ies2}).

The effective Hamiltonian induced by (\ref{newint2}), contributing to the $B^0_d\leftrightarrow
\bar{B}^0_d$ transitions is:
\begin{eqnarray}
\mathcal{H}^{\Delta B=2}_{eff}\vert_{h+A}&=&\sum_i\frac{1}{8 m^2_i}[(I^{i+}_B)^2_{bq}(\bar{b}q)^2
+(I^{i-}_B)^2_{bq}(\bar{b}\gamma_5q)^2]
\nonumber \\&-&
\sum_i\frac{1}{8 m^2_A}[[(I^{i+}_B)^A_{bq}]^2(\bar{b}q)^2+[(I^{i-}_B)^A_{bq}]^2(\bar{b}\gamma_5q)^2].
\label{heff4}
\end{eqnarray}
and, as usual we define
\begin{equation}
\Delta M_{B_d}\vert_{h,A}= 2\textrm{Re}\langle\bar{B}^0\vert\mathcal{H}^{\Delta B=2}_{eff}\vert_{h}
\vert B^0\rangle=\textrm{Re}\zeta^h_{bd}\,
 \langle \bar{B}^0\vert (\bar{b}d)^2_{V-A} \vert B^0\rangle
\label{effective4}
\end{equation}
where
\begin{eqnarray}
\textrm{Re}\,\zeta^h_{bd}=\textrm{Re}\sum_i\frac{1}{32 m^2_i}\left[0.6(I^{i+}_{B_d})^2_{bd}
-16.5(I^{i-}_{B_d})^2_{bd} \right]\,\textrm{GeV}^{-2},
\label{zetas4}
\end{eqnarray}
and
\begin{equation}
(I^{i\pm}_{B_d})^2_{bd} =  [(U_{\rho i}-0.2U^*_{\eta i})U_{\rho i}\pm 0.2U_{\eta i}U_{\rho i}] \times 10^{-8},
\end{equation}
then
\begin{eqnarray}
&&\textrm{Re}\,\zeta^h_{bd}=\textrm{Re} \sum_i \frac{1}{32 m_i^2} \{0.6(
(U_{\rho i}-0.2U^*_{\eta i})U_{\rho i}+ 0.2U_{\eta i}U_{\rho i}
)) \nonumber \\&&- 16.5((U_{\rho i}-0.2U^*_{\eta i})U_{\rho i}- 0.2U_{\eta i}U_{\rho i}) \}\times10^{-8}.
\label{zetas5}
\end{eqnarray}

Assuming that only one of the scalars contribute in (\ref{zetas2}), we obtain a constraint on the contributions
of $Z^\prime$, one scalar
$h^0_1$ and one pseudoscalar $A^0_1$ to $\Delta M_B$, like that in Eq.~(\ref{ok}) for the kaon system:
\begin{eqnarray}
&& 10^{-1}\,
\frac{M^2_W}{M^2_{Z^\prime}}\,+ \frac{1}{32 m^2_1} \{0.6[
(U_{\rho 1}-0.2U^*_{\eta 1})U_{\rho 1}+ 0.2U_{\eta 1}U_{\rho 1}] \nonumber \\ && - 16.5
[(U_{\rho 1}-0.2U^*_{\eta 1})U_{\rho 1}- 0.2U_{\eta 1}U_{\rho 1}]-\mathcal{A}^\prime
\} \,< 10^{-4} \,\textrm{GeV}^{-2}.
\label{ok2}
\end{eqnarray}
where $\mathcal{A}^\prime$ is the pseudoscalar contribution which is also similar to that of the scalar one
in (\ref{zetas5})
but with $m_1\to m_A$ and $U_{\eta 1}\to V_{\eta 1}$ and $U_{\rho 1}\to V_{\rho 1}$. The analysis of the
$B_s-\bar{B}_s$ system follows the same
procedure.

As can be seen from Figs.~\ref{fig1},\ref{fig2} and \ref{fig4}, the constraints coming from $B_d-\bar{B}_d$ are
stronger than those in the $K^0-\bar{K}^0$ and $B_s-\bar{B}_s$. Moreover, in the $B_d$ system  the interference
of $Z^\prime$ with the pseudoscalar is what matters, although this is not as important as in the kaon system.
See Sec.~\ref{sec:results} for discussions. We will see in the next section that the interference
is more dramatic in the $\Delta M_D$ case.


\section{What Higgs boson is this?}
\label{sec:whathiggs}

We have assumed that the mass of the lightest scalar is equal to that of the resonance found at
LHC~\cite{cmsatlas}.
We see from Fig.~\ref{fig1} that the values of the $M_{Z^\prime}$ allowed by $\Delta F=2$ processes
depend on $U_{\rho 1}$ and
$U_{\eta 1}$  matrix elements in the neutral scalar sector. The other factor denoted by $\mathcal{\bar{K}}^{Ux}_{q_1q_2}$,
have already been fixed.  Assuming that the production processes are
the same of the SM (new
sources should be suppressed by the masses of the extra particles of the model), the neutral scalar $h^0_1$,
must couple to fermions, at least to the top quark,  with a similar strength to that in the SM, in order to
have a compatible Higgs boson production rate.
The latter point is important since the new resonance discovery at LHC~\cite{cmsatlas} is still compatible with the SM expectation and it
has couplings to fermion and vector bosons compatible with the
SM Higgs~\cite{giardino}. In $d$-type quark sector we have already seen that only the $b$-quark has a
coupling to that resonance that can be smaller than the SM one.

From Eqs.~(\ref{yukan}) and (\ref{yukanma}), the $u$-type quark-neutral-scalar couplings are
\begin{eqnarray}
&&\mathcal{K}^U\approx\left(
\begin{array}{ccc}
0.0099\rho^0 -10^{-6}\eta^0 & 0.00340\rho^0-10^{-5}\eta^0 & 0.0109\rho^0-10^{-5}\eta^0\\
-0.13846\rho^0+10^{-7}\eta^0&0.0556\rho^0+10^{-6}\eta^0 &-0.1521\rho^0-10^{-6}\eta^0\\
1.9228\rho^0-10^{-11}\eta^0&0.8656\rho^0-10^{-10}\eta^0&2.3569\rho^0-10^{-10}\eta^0
\end{array}
\right).
\label{kapasu}
\end{eqnarray}

\subsection{$\Delta M_D$ }
\label{subsec:mc}

Let us now consider a $\Delta C=2$ process: the mass difference between charmed neutral mesons, $\Delta M_D$.
We use the numbers in Eqs.~(\ref{numeros}) and (\ref{kapasu}) for the transition $D^0\leftrightarrow \bar{D}^0$.
$(K^U_L)_{uc}\sim 8.9\times10^{-2}$, and $(I^{1\pm})_{uc}=[0.0034\mp 0.138]U_{\rho1}$, that is $(I^{1+})_{uc}=-0.057$  and
$(I^{1-})_{uc}=0.06$. Hence, We obtain
\begin{equation}
\Delta M_D=\left( 1.3\frac{M^2_W}{M^2_{Z^\prime}}-4.05\right)\times10^{-9}\;\textrm{GeV},
\label{nova1}
\end{equation}
and we see that $M_{Z^\prime}>43$ GeV gives agreement with the experimental value already, but with the $Z^\prime$ alone we would
obtain $M_{Z^\prime}>27$ TeV.

\subsection{Higgs-$u$-quark couplings}
\label{subsec:hf}

The Yukawa couplings in the SM are $y_u=1.3\times10^{-5}$, $y_c=7.3\times10^{-3}$ and $y_t=0.997$.
In the present model these values correspond to the diagonal entries in the matrix (\ref{kapasu}). From the latter,
we see that the couplings of the $u,c$-quarks are dominated by the neutral scalar $\rho^0$, and it may be compatible
with the SM values depending of the values of $U_{\rho1}$. From Eq.~(\ref{kapasu}) we see
that the larger coupling of $h^0_1$ is with the top quark and can be numerically equal to the coupling in the
SM, if $U_{\rho 1}=0.42$ regardless the value of $U_{\eta 1}$, i.e., $(\mathcal{\bar{K}}^{U\rho})_{tt}U_{\rho1}=
\sqrt{2}m_t/v\sim0.9974$. In this case we have that $(\mathcal{\bar{K}}^{U\rho})_{uu}U_{\rho1}=0.0042$ and
$(\mathcal{\bar{K}}^{U\rho})_{cc}U_{\rho1}=0.0233$. These values are larger than the respective ones in the
SM. However this is not a problem by now. With the present data it is not possible to measure $y_c$ directly.

Nevertheless this may not be the full history. In 3-3-1 models there are extra heavy quarks, and hence, the
gluon fusion can produce the SM-like Higgs throughout new diagrams involving the extra quarks. These exotic quarks
are singlet non-quiral quarks under the gauge symmetry of the SM. However, they are quiral quarks under the 3-3-1
symmetry and couple to the neutral scalar $\chi^0$, which is a singlet under the SM gauge symmetries and
has a projection on the SM-like neutral scalar given by $U_{\chi 1}$. Hence, $gg\to h^0_1$ may have contributions
from these exotic quarks that are proportional to $U^2_{\chi 1}$, but independent of the exotic quark masses,
they would be smaller than $U_{\eta1}$, and $U_ {\rho1}$ since the $\chi^0$  must have its main projection on a heavy
neutral scalar. These parameters together can still mimic  the SM Higgs production unless the exotic quarks are
too heavy, or $U_{\chi1}$ is very small, as we are assuming here. However, if the exotic quarks are not too
heavy, or $U_{\chi1}$ is larger than we thought, these quarks will contribute significantly to the $h^0_1$
production but, since at the same time the rates will be reduced, it is possible that some observables do not
change. In the latter case we could consider $U_{\rho 1}$ again as a free parameter and the $h^0_1$ does not
necessarily is the SM-like Higgs. It can be one of the extra Higgs in the model i.e., it is not the resonance that
was discovered at LHC.

As we said before, the couplings of the 125 GeV Higgs boson to $W$, $Z$ and fermions may have strength
that can be smaller than the respective couplings in the SM~\cite{gupta} since these couplings are modify by the
matrix elements like $U_{\eta1}$ and $U_{\rho1}$. However, the Yukawa couplings, see as in Eqs.~(\ref{kapasd}) and
(\ref{kapasu}), may be larger or smaller than the SM couplings~\cite{flavour}.
For instance, top quark decay $t\to c\,h^0_1$ is now possible, and written the respective couplings by
$\bar{c}(a+i\gamma_5b)t$, we have from Eq.~(\ref{ok2}) that $a=[(\mathcal{K}^U)_{ct}+
(\mathcal{K}^U)^*_{tc}](U_{\rho1}/2)$ and $b=[(\mathcal{K}^U)_{ct}-(\mathcal{K}^U)^*_{tc}](U_{\rho1}/2)$. Using
the numerical values in Eq.~(\ref{ok2}) we see that $a\approx0.15$ and $b\approx0.21$. The value of $(a+b)/2
\approx0.18$ may be considered consistent with the recent upper limit for the coupling of the vertex $\bar{c}th^0_1$
obtained by the ATLAS: $a<0.17$~\cite{atlasnote}. Having fixed the values of $U_{\rho1}$ and allowing $U_{\eta1}$ to
run over the range $[-0.2,0.2]$ and the other numbers in Eq.~(\ref{kapasu}) we do not have any freedom with the $\Delta M_D$
observable. In this case the interference between the neutral scalar and the $Z^\prime$ contributions are more dramatic:
only the $Z^\prime$ implies $M_{Z^\prime}>27$ TeV and the scalar contributions alone give a large contribution:
both, however, imply $M_{Z^\prime}>42.5$ GeV.  Here we have not consider the pseudoscalar contributions to $\Delta M_D$.
Once again we would like to emphasize that all of this is at $\mu=M_Z$.

In the next section we consider the $\vert\Delta F\vert=1$ forbidden processes.

\section{$\Delta F=1$ processes}
\label{sec:deltaf1}

Concerning the $\vert\Delta F\vert=1,\;F=S,B$ processes,  we consider as an illustration the
leptonic decays of  neutral mesons, $M^0$, i.e.,
$M^0\to l^+l^{\prime -}$, with $l,l^\prime=e,\mu$; and $M^0$ an strange or a beauty meson.
We recall that these processes, at the tree level, involve only one vertex in the quark sector and the $Z^\prime$
has natural flavor conservation in the lepton sector. When a (pseudo)scalar is exchanged, the other vertex involves
the interactions of the charged leptons that do not conserve the lepton flavor. This is parameterized by the arbitrary
matrix $U_{ll^\prime}$ as discussed below.

In the m3-3-1 model the partial width of the decay $M^0(q_1\bar{q}_2)\to l ^+l^{\prime -}$,
where $M^0=K,B_{d,s}$ has contributions at tree level
which, are given by
\begin{eqnarray}
\mathcal{B}^{331}_{M\to l^+l^{\prime -}}\!&=&\!\left\{\frac{G_FM^2_W}{16\sqrt{2}c^2_W}
\vert (K^D_L)_{q_1q_2}\vert^2\frac{f^2_M M^2_Mm^2_l}{M^4_{Z^\prime}} \!
+\!
\frac{M^6_Mf^2_M}{2(m_{q_1}\!+\!m_{q_2})^2}\left[\left\vert
\frac{(I_M)_{q_1q_2}U_{ll^\prime}}{m^4_{h_1}}\right\vert^2+
\left\vert\frac{(I_M)^A_{q_1q_2}U^A_{ll^\prime}}{m^4_A}\right\vert^2 \right]\right.
\nonumber \\\!&-&\!\left.
\frac{(\sqrt{2}G_FM^2_W)^{\frac{1}{2}}}{64c_W}\frac{M^4_M f^2_M m_l}{(m_{q_1}\!+
\!m_{q_2})M^2_{Z^\prime}M^2_{h_1}}(K^D_L)_{q_1q_2}(I_M)^*_{q_1q_2}
U^*_{ll^\prime}\right\}
\frac{(1\!-\!\frac{4m^2_l}{M^2_M})^{\frac{3}{2}}\tau_M}{16\pi M_M},
\label{gamma331}
\end{eqnarray}
where $\tau_M$ is the meson $M^0$ half-life, $M_M$ its mass and we have used the meson matrix
elements
\begin{equation}
\langle0\vert\bar{q}_f\gamma_\mu\gamma_5 q_i\vert M^0\rangle=if_Mp^\mu_M,\quad \langle0
\vert\bar{q}_f\gamma_5 q_i\vert M^0\rangle=-if_M\,
\frac{M^2_M}{m_{q_f}+m_{q_i}},
\label{pcac}
\end{equation}
and $p_M=p_1+p_2$.

The matrix $U_{ll^\prime}$ in (\ref{gamma331}) arises as follows. The three lepton generations
transform under the 3-3-1 symmetry as
$\Psi_a=(\nu_a\,l_a\, l^c)^T_L\sim (1,\textbf{3},0)$ and we do not introduce right-handed neutrinos.
The Yukawa interactions in the lepton sector are:
\begin{equation}
-\mathcal{L}_{\nu l H}=\epsilon_{ijk}\overline{(\Psi_{ia})^c}G^\eta_{ab}\Psi_{jb}\eta_k+
\overline{(\Psi_{ia})^c}G^S_{ab}\Psi_{jb}S^*_{jk}+H.c.,
\label{leptons}
\end{equation}
where $a,b$ are generations indices, $i,j,k$ are $SU(3)$ indices, and $G^\eta$ ($G^S)$ is a
antisymmetric (symmetric) matrix.
In Eq.~(\ref{leptons}), $\eta$ is the same triplet which couples to quarks and $S$ is a sextet,
$S\sim(1,\textbf{6},0)$ which does not couple to quarks. Under $SU(2)_L\otimes U(1)_Y$ the sextet
transform as $S=\textbf{1}+\textbf{2}+\textbf{3}$,
and we see that there is a doublet and a non-Hermitian triplet which gives mass to charged leptons
and active left-handed neutrinos, respectively.
However, although the sextet is enough  to give to neutrinos a Majorana mass and a Dirac mass to
the charged leptons, it does not give a $PMNS$ matrix
$V_{PMNS}=U^{l\dagger }_LU^\nu_L$, since when only the sextet is the source of lepton masses we
have that
$U^l_L=U^\nu$. Hence the interaction with the $\eta$ triplet is mandatory. In this case the mass
matrices of the neutrinos and charge leptons are
\begin{equation}
-\mathcal{L}^\nu_M=\overline{(\nu_{aL})^c}G^S_{ab}\nu_{bL}\frac{v_{\sigma_1}}{\sqrt{2}}+H.c.,\quad
\mathcal{L}^l_M=
\bar{l}_{iaL}\left[G^\eta_{ab}\frac{v_\eta}{\sqrt2}+
G^S_{ab}\frac{v_{\sigma_2}}{\sqrt2}\right]l_{jbR}+H.c.,
\label{leptons2}
\end{equation}
where $\sigma^0_1$ and $\sigma^0_2$ are the neutral component of the triplet and doublet in the
sextet, respectively. In terms of the mass eigenstates
we have $Re\,\sigma^0_1=\sum_iU_{Si}h^0_i$,
$Re\,\sigma^0_2=\sum_i U_{Di}h^0_i$, $Im\,\sigma^0_1=\sum_iV_{Si}h^0_i$ and $Im\,\sigma^0_2=
\sum_i V_{Di}h^0_i$.

We have $M^\nu_{ab}=G^S_{ab}\frac{v_{\sigma_1}}{\sqrt{2}}$ and
$M^l_{ab}=G^\eta_{ab}\frac{v_\eta}{\sqrt2}+G^S_{ab}\frac{v_{\sigma_2}}{\sqrt2}$.
These mass matrices are diagonalized as follows: $\hat{M}^\nu=U^{\nu T}_LM^\nu U^\nu_L$ and
$\hat{M}^l=U^l_L M^l U^{\dagger}_R$ and the relation
between symmetry eigenstates (primed) and mass (unprimed) fields are $l^\prime_{L,R}=U^l_{L,R}l_{L,R}$
and $\nu^\prime_L=U^\nu_L \nu_L$, where
$l^\prime_{L,R}=(e^\prime,\mu^\prime,\tau^\prime)^T_{L,R}$, $l_{L,R}=(e,\mu,\tau)^T_{L,R}$  and
$\nu ^\prime_L=(\nu_e\,\nu_\mu\,\nu_\tau)^T_L$
and $\nu_L=(\nu_1\,\nu_2\,\nu_3)_L$.

The interactions of the neutral scalars and pseudo-scalars with the leptons are
\begin{eqnarray}
-\mathcal{L}_{leptons}&=&\sum_{i,n=1,2,3}\overline{(\nu_{nL})^c}\nu_{nL}
\frac{\sqrt2 m_{\nu_n}}{v_{\sigma_1}}(U_{S i}h^0_i+iV_{S i}A^0_i)\nonumber \\ &+&
\sum_{i,l,l^\prime}\bar{l}_{L}U^{l\dagger}_L\{[ G^\eta_{ll^\prime}U_{ \eta i} +
G^S_{ll^\prime}U_{Si}]h^0_i+  i [G^\eta_{ll^\prime}V_{\eta i}+
G^S_{ll^\prime}V_{D i}]A^0_i)\}U^l_Rl^\prime_{ R}+H.c.
\label{leptons3}
\end{eqnarray}
where $l,l^\prime=e,\mu,\tau$. For one scalar $h^0_1$ and one pseudoscalar $A^0_1$ we have
\begin{equation}
U_{ll^\prime}= U^{l\dagger}_L[G^\eta_{ll^\prime}U_{\eta 1} +G^S_{ll^\prime} U_{D 1}]U^l_R,\quad
U^A_{ll^\prime}=U^{l\dagger}_L[G^\eta_{ll^\prime}V_{\eta 1}+ G^S_{ll^\prime}V_{D 1}]U^l_R,
\label{leptons4}
\end{equation}
respectively. To be consistent with our previous analysis $U_{D1}$ and $V_{D1}$ have to be smaller
than the other entries of the $U$ and $V$ matrices and
can be neglected. It is the arbitrary matrix in Eq.~(\ref{leptons4}) which appears in Eq.~(\ref{gamma331}).
Notice that it is sum of two products involving four
matrices each. The FCNC effects in the charged lepton sector can be avoided only by fine tuning as
$G^\eta_{e\mu}\frac{v_\eta}{2}+G^S_{e\mu}
\frac{v_{\sigma_2}}{2}=0$, etc. Otherwise, we have processes like $l\to l^\prime\gamma$,
$l\to \bar{l}^\prime l^\prime l^{\prime\prime}$ , where
$l=\mu,\tau$ and $l^\prime,l^{\prime\prime}=e,\mu$. For instance, experimentally it is found
$\mathcal{B}_{\mu\to e+\gamma}<5.7\times10^{-13}$ \cite{meg}, a
value still well above the SM prediction $\sim10^{-52}$~\cite{cheng}.  In the present model this decay occurs at
the 1-loop level too. On the other hand decays like $\mu^+\to e^+e^-e^+$. with branching ratio
$<10^{12}$~\cite{sindrum} occurs at the tree level mediated by neutral scalars, in particular by the $h^0_1$.
The branching ratio of this decay in the m3-3-1 model is proportional to $(1/G^2_Fm^4_1)\vert U_{ee}U_{e\mu}\vert^2\sim3
\times10^{-3}\vert U_{e\mu}\vert^2$ and constrains mainly the non-diagonal matrix element
$U_{e\mu}$ which we recall is the arbitrary matrix defined in Eq.~(\ref{leptons4}). Since
$U_{ee}<10^{-2}$ with the larger values corresponding to the case when we consider also the pseudoscalar
(see Fig.~\ref{fig11}), we have that $\vert U_{e\mu}\vert^2<10^{-5}$. We can see from the definition of the $U$ matrix
in Eq.~(\ref{leptons4}) that it is not a too-strong constraint since this matrix is the sum of the two products of
four matrices with two of them being ($G^\eta,G^S$) arbitrary ones. Decays like $h^0_1\to \mu^+\tau^-$ can be observable at the
LHC~\cite{blankenburg}. More details on this will appear elsewhere.

It worth to call to the attention that the m3-3-1 does not need the introduction of singlet
right-handed neutrinos for having massive (light) active Majorana neutrinos and also accommodated the PMNS mixing matrix.
If we add right-handed neutrinos and avoid, by an appropriate symmetry, the coupling of $\eta$ to leptons we have
$M^\nu_{ab}=G^\eta_{ab}\frac{v_\eta}{\sqrt2}+ G^S_{ab}\frac{v_{\sigma_1}}{\sqrt{2}}$ and
$M^l_{ab}=G^S_{ab}\frac{v_{\sigma_2}}{\sqrt2}$ and the FCNC arise in the neutrino sector. I
n the most general case FCNC occur in both sectors. The 3-3-1 model with right-handed neutrinos transforming non-trivially
under $SU(3)_L$ was first put forward by Montero \textit{et al}., in Ref.~\cite{mpp}. Anyway, if sterile right-handed
neutrinos (with respect to the SM interactions) do exist, they can be accommodated in an
$SU(4)_L\otimes U(1)_{X^\prime}$ model, see Ref.~\cite{pp94}. Summarizing, the m3-3-1 model
ought to have FCNC in the scalar-charged lepton in\-te\-rac\-tions if no right-handed neutrinos (transforming as singlets
under $SU(3)_L$) are added to the matter content of the model.

Now, we are able to discuss leptonic decay of neutral mesons.

\subsection{$K_L\to l^+l^-$}
\label{subsec:kaondecay}

The experimental data are $\mathcal{B}_{K_L\to e^+e^-}<10^{-12}$ and
$\mathcal{B}_{K_L\to \mu^+\mu^-}=(6.84\pm0.11)\times 10^{-9}$~\cite{pdg}.
Using $q_1=s$ and $q_2=d$, $M_M=M_K$, $f_M=f_K$, we obtain from Eq.~(\ref{gamma331}) that
the decay into electrons imposes a strong bond on the
values of $U_{\eta 1}$ but not on $M_ {Z^\prime}$. This is shown in Figs.~\ref{fig5}  for
$K_L\to e^+e^-$ decay.
We have an additional free parameter, $U_{ee}$, that weakens this bonds, see Fig.~\ref{fig6}.
For the $K_L\to \mu^+\mu^-$ decay see Fig.~\ref{fig7}.
On the other hand, the bound from the two muon decay on the $Z^\prime$ mass it is less restrictive
than $K_L\to e^+e^-$. See also the discussion
in Sec.~\ref{sec:results}.

\subsection{$B_{s,d} \to \mu^+\mu^-$}
\label{subsec:bdecay}
Next we consider $\Delta B=1$ processes. Recently,
it has been observed the branching ratio $\mathcal{B}_{B^0_s\to \mu^+\mu^-}=3.2\times 10^{-9}$
and $\mathcal{B}_{B^0_d\to \mu^+\mu^-}=8 \times
10^{-10}$~\cite{lhcb}. In Both cases there is not constraint in $U_{\mu\mu}$, however, $U_{\eta 1}$
has the biggest constraint from the
$\mathcal{B}_{B^0_d\to \mu^+\mu^-}$ decay, as can be seen in Fig.~\ref{fig8} (the solid gray curve),
being the allowed interval around $[-0.5,0.5]$.
The constraint on $M_{Z^\prime}$, for both cases, is weaker than those coming from the other processes.
In fact, these decays allow a
rather light $Z^\prime$ as is shown in Fig.~\ref{fig9}. In the latter figure we show also the constraint
coming from the $K_L$ decays and $\Delta M_K$.

\section{Results}
\label{sec:results}

Here we will discuss in more details the constraints on the $Z^\prime$ mass taking into account both the
scalar and pseudoscalar contributions to
the processes discussed above. First, let us consider the $\Delta M_M,\;M=K,B_{d,s}$ cases. In each case
we first consider only the scalar $h^0_1$
contribution by considering the pseudoscalar very heavy ($m_A\to \infty$), in practice we have used
$m_A=100$ TeV when we want to decouple the
pseudoscalar $A^0_1$ from Eqs.~(\ref{ok}), (\ref{ok2}) and (\ref{gamma331}). For the sake of simplicity
we consider that the mixing matrix elements
in the scalar and pseudoscalar sector have the same numerical values, i..e, $U_{\eta1}=V_{\eta1}$ and
$U_{\rho1}=V_{\rho1}$. We are also assuming
that the other scalars and pseudoscalars in the model, even if their projections on $h^0_1,A^0_1$ are
large, are heavy enough to give no observable
effects in the processes considered above. It implies that even if we use $U_{\rho1}=V_{\rho1}=0.42$,
$U_{\eta1}$ and $V_{\eta1}$ are still free parameters
but we will consider them to be equal, i..e, $U_{\eta1}=V_{\eta1}$, just for simplicity.
We would like to recall that all our results  are consequences of the mixing matrices $V^{U,D}_{L,R}$
obtained in Sec.~\ref{sec:331}.

\begin{figure}[ht]

\begin{center}

\includegraphics[width=5in]{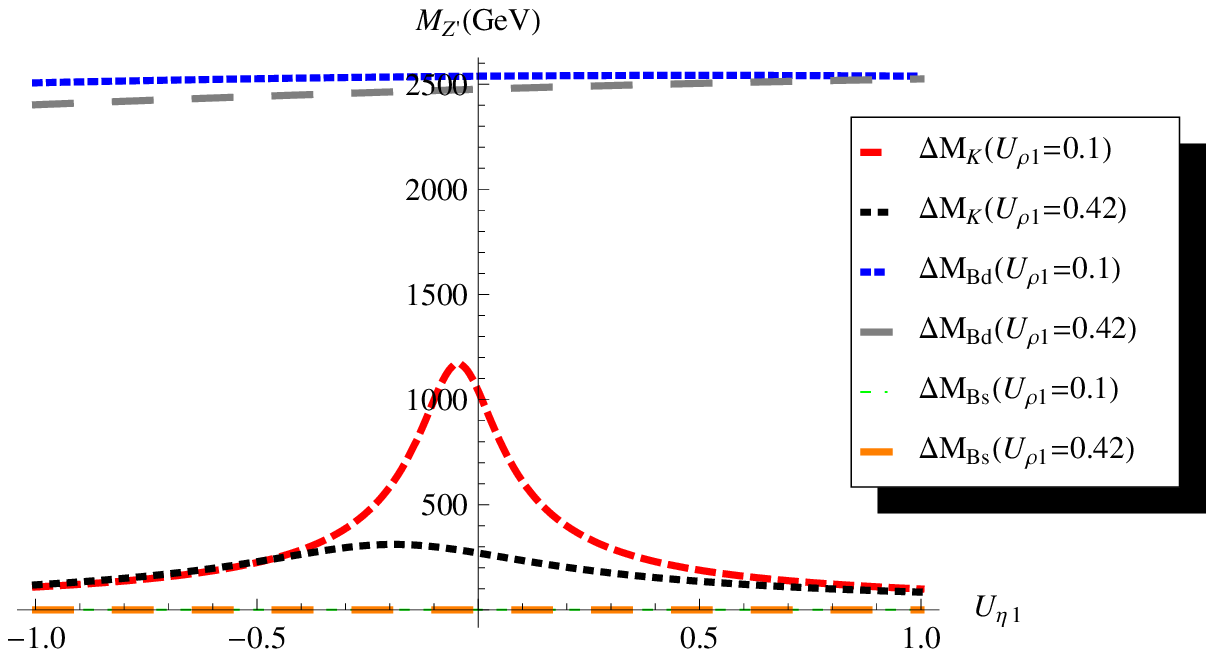}
\caption{$Z^\prime$ mass values  satisfying  Eq. (\ref{ok}) and Eq. (\ref{ok2}), simultaneously, but not including the pseudoscalar contribution,
for a fixed value of the element $U_{\rho 1}$ ($U_{\eta 1}$), and the other  $U_{\eta 1}$ ($U_{\rho 1}$) running in the range
$[- 1, 1]$.}
\label{fig1}
\end{center}
\end{figure}

\begin{figure}[ht]
\begin{center}
\includegraphics[width=5in]{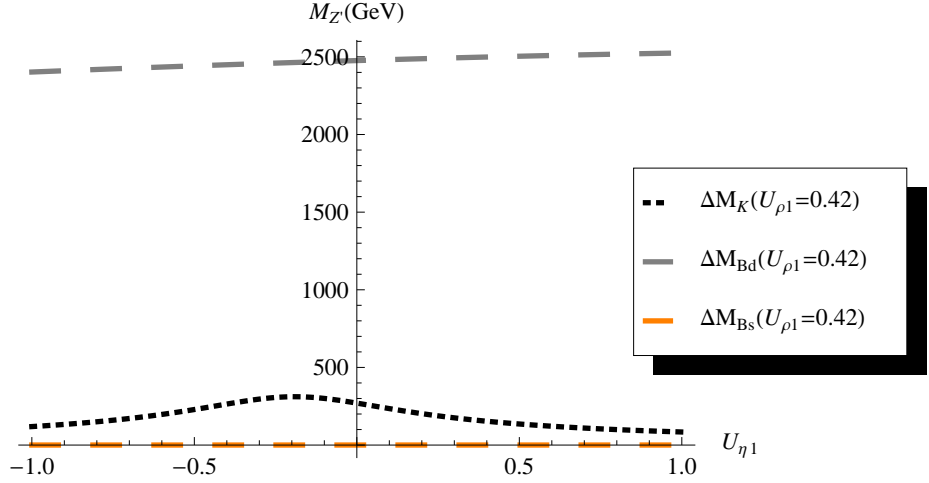}
\caption{Same as Fig.~\ref{fig1} but now with
$U_{\rho 1} = 0.42$ (the value that ensures that the coupling of $ h_1^0 $ with
the top quark is equal to the SM), and  $U_{\eta 1}$ running in the interval
$[- 1, 1]$. }
\label{fig2}
\end{center}
\end{figure}

\begin{figure}[ht]
\begin{center}
\includegraphics[width=6in]{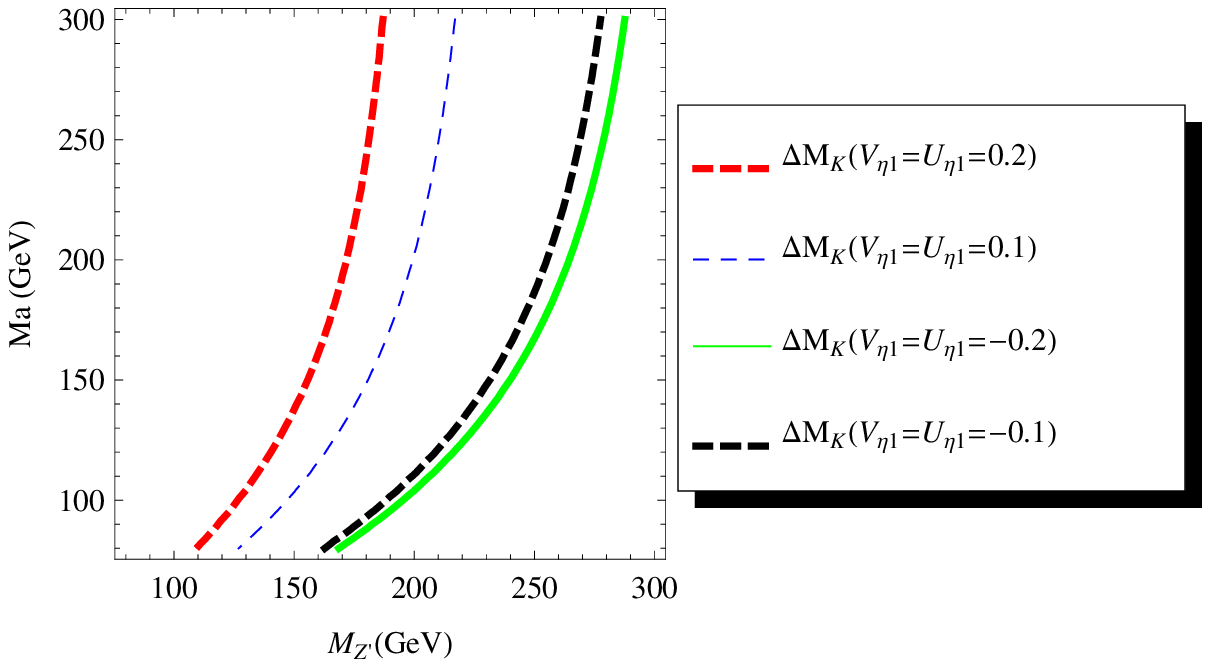}
\caption{ Considering  Eq. (\ref{ok}) with the contribution of the pseudoscalar. The allowed region for the $Z^\prime$ mass and
the pseudoscalar mass $M_a$ were obtained by setting values for $U_{\eta 1}=V_{\eta 1}$. The smallest value to the
$Z^\prime$ mass is when $U_{\eta 1} = V_{\eta 1} = 0.2$ and the biggest  when $U_{\eta 1} =V_{\eta1}= - 0.2$.}
\label{fig3} 
\end{center}
\end{figure}

\begin{figure}[ht]
\begin{center}
\includegraphics[width=6in]{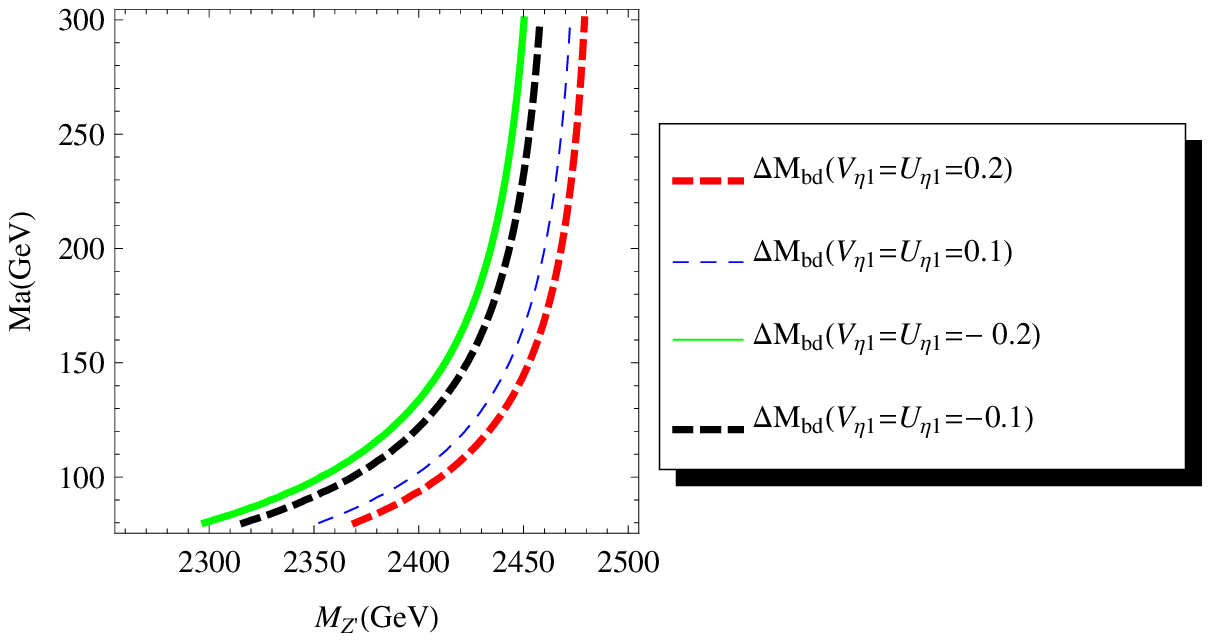}
\caption{Considering the Eq. (\ref{ok2}) with the contributions of the pseudoscalar.
The allowed region for the $Z^\prime$ mass and
the pseudoscalar mass $M_a$ were obtained by setting values for $U_{\eta 1}=V_{\eta 1}$. The smallest value to the
$Z^\prime$ mass is when $U_{\eta 1} = V_{\eta 1} = -0.2$ and the biggest  when $U_{\eta 1} =V_{\eta1}= 0.2$.}
\label{fig4} 
\end{center}
\end{figure}

The scalar contribution (when the pseudoscalar is considered too heavy) to the $\Delta M_M$ mass
differences are shown in Fig.~\ref{fig1}. In this figure we show the values of $M_{Z^\prime}$, as a
function of $U_{\rho1}$ ($U_{\eta1}$) for fixed $U_{\eta1}$ ($U_{\rho1}$), which are allowed by solving
simultaneously Eqs. (\ref{ok}) and (\ref{ok2}). In principle, both $U_{\rho1}$ and $U_{\eta1}$ are allowed
to vary in the interval $(-1,1)$. We see from this figure that a large range for the $Z^\prime$ mass values
is allowed by $K$-mesons but not by the $B_s$ and $B_d$ mesons. Notice also that under our conditions in
Sec.~\ref{sec:331}, $\Delta M_{B_s}$ does not constrain $m_{Z^\prime}$ at all. However, $\Delta M_{B_d}$
does: $m_{Z^\prime}>2.5$ TeV. On the other hand, by demanding that $\rho^0$ be equivalent
to the SM Higgs implies, from Eq.~(\ref{kapasu}), $U_{\rho1}=0.42$, see Sec.~\ref{sec:whathiggs}. In this case,
the only variable is $U_{\eta1}$, and the $Z^\prime$ mass can still be of the order of the electroweak
scale or even lower. Fig.~\ref{fig2} shown the same as Fig.~\ref{fig1} but now with $U_{\rho1}=0.42$. There
is negative interference in the $K$ mesons system between the $Z^\prime$ and $h^0_1$
amplitudes: without the scalar contribution $\Delta M_K$  implies also $m_{Z^\prime}>2.5$ TeV. In the
$B_{d,s}$ systems the interference is not important. If we allow the $A^0_1$'s mass to be a free parameter,
we show in Figs.~\ref{fig3} and \ref{fig4} the effects of this pseudoscalar in the $K$ and $B_{d,s}$ systems.
Those figures show the allowed values for the masses $m_A$ and $m_{Z^\prime}$.
For obtaining Figs.~\ref{fig3} and \ref{fig4} we
have assumed that $U_{\eta1}=V_{\eta1}$ and $U_{\rho1}=V_{\rho1}=0.42$. Notice that now there is
negative interference between $Z^\prime$ and the
pseudoscalar, $A^0_1$, implying a smaller lower bound on $m_{Z^\prime}$ in the $B_d$ system:
$m_{Z^\prime}>2.3$ TeV. In the case os the $B_s$ system
the scalar and pseudoscalar are not important and the constraint on the $Z^\prime$ mass is weaker
than in the other mesons.

\begin{figure}[ht]
\begin{center}
\includegraphics[width=15cm,height=12cm,keepaspectratio]{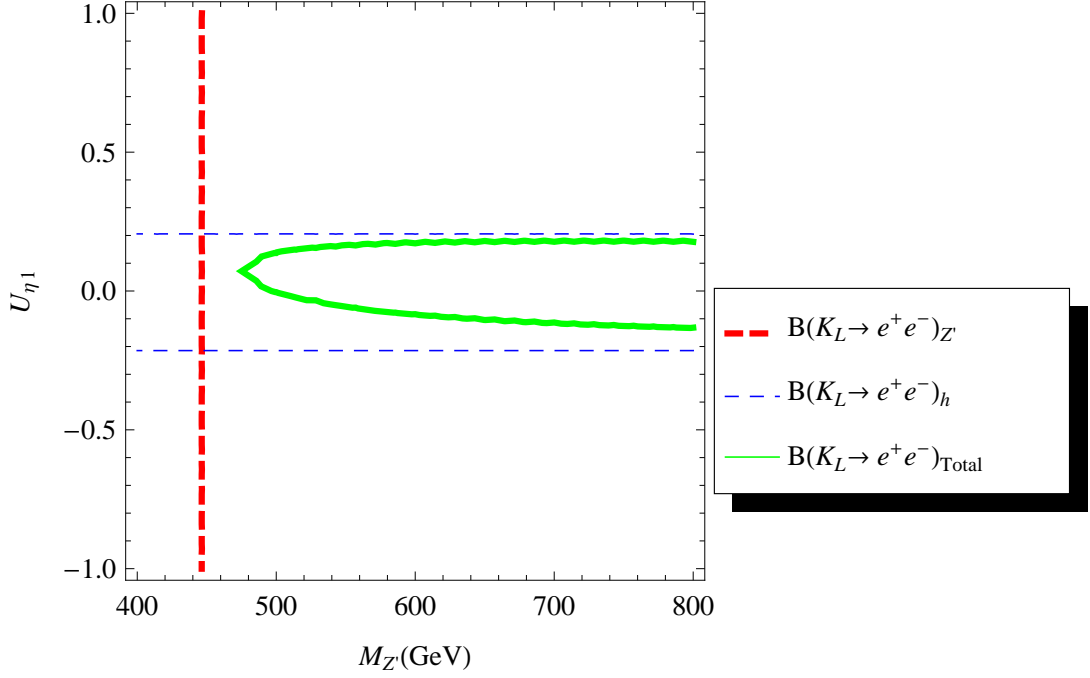}
\caption{Allowed region for $M_{Z^\prime}$ and $U_{\eta1}$, for fixed $U_{ee}=10^{-4}$ by the $K_L\to ee$ decay, using the Eq.~(\ref{gamma331}),
with $l=l^\prime=e$ and $M=K$). The red (dashed) vertical line is the contribution of $Z^\prime$ only, and the allowed range is to the right of the curve.
The region within the blue (dashed) horizontal lines is the allowed region for the scalar contribution only.
The total contribution is given by the green (continuous) curve and the allowed region is the area within this curve. Notice that we are not considering
the pseudoscalar yet.}
\label{fig5}
\end{center}
\end{figure}

\begin{figure}[ht]
\begin{center}
\includegraphics[width=5in]{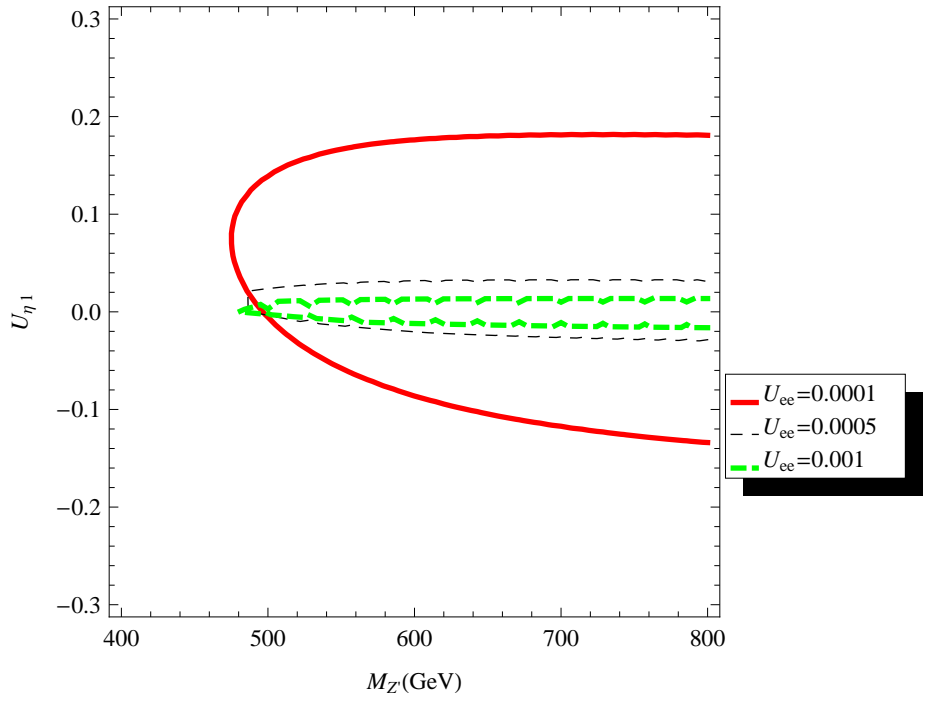}
\caption{Same as Fig.~\ref{fig5} but now showing the dependence on $U_{ee}$. The red (continuos) line is with $U_{ee}=10^{-4}$, $U_{ee}=5 \times 10^{-4}$ black
(thin dashed) line and $U_{ee}=10^{-3}$ green (thick green) line. The allowed region is always to the right and bounded by the curves.}
\label{fig6}
\end{center}
\end{figure}

From Fig.~\ref{fig5} we see that in the case of $K_L\to e^+ e^-$ decay, the interference between
$Z^\prime$ and $h^0_1$ is constructive, assuming the contribution of the $A^0_1$ negligible. We use
Eq.~(\ref{gamma331}) with $l=l^\prime=e$ and $M^0=K$.  The value of the $M_{Z^\prime}$ mass change,
from 440 GeV when only the $Z^\prime$ contribution is consider, to 460 GeV when both the $Z^\prime$ and $h^0_1$
contribution are taken into account. The pseudoscalar contribution is omitted in this figure. The figure
shows the allowed values for $M_{Z^\prime}$ and $U_{\eta1}$, for fixed $U_{ee}=10^{-4}$, see Eq.~(\ref{leptons4}),
by this decay. The red (dashed) vertical line is the contribution of $Z^\prime$ only and the allowed range is
to the right of the curve. The minimal value allowed by this decay is around 445 GeV. The blue (dashed)
horizontal lines are the contributions of the scalar only and the allowed range for $U_{\eta1}$,
i.e., $-0.2< U_{\eta1}<0.2$ for any value of $M_{Z^\prime}$. The total contribution is given by the green
(continuous) curve and the allowed region
is inside that curve, and the minimal value for $M_{Z^\prime}$ has moved to 500 GeV. Fig.~\ref{fig6} shows
the total contribution (the green curve
in Fig.~\ref{fig5}) for several values of $U_{ee}$. Notice that $U_{\eta1}$ is constrained, $\vert U_{\eta1}\vert<0.2$.

\begin{figure}[ht]
\begin{center}
\includegraphics[width=15cm,height=12cm,keepaspectratio]{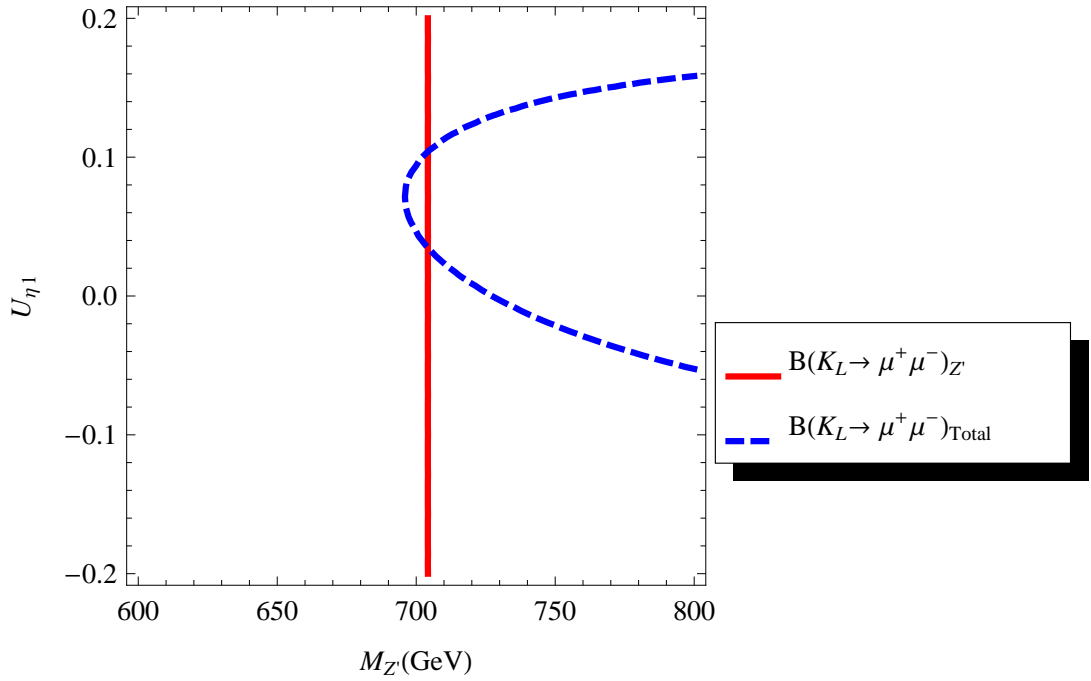}
\caption{Constraints on $M_{Z^\prime}$ and $U_{\eta1}$  from the $K_L\to \mu \mu$ decay fixed $U_{\mu \mu} = 0.01$, using Eq.~(\ref{gamma331}) with $l=l^\prime= \mu$ and $M=K$.
The allowed regions are those to the right of  the curves, for $M_{Z^\prime}$ and $U_{\eta1}$.  The red (solid) vertical line is the  contributions of the $Z^\prime$
only. The blue (dashed) curved line is the total contribution to the decay. We consider only the $Z^\prime$ and the scalar contributions.}
\label{fig7}
\end{center}
\end{figure}

\begin{figure}[ht]
\begin{center}
\includegraphics[width=6in]{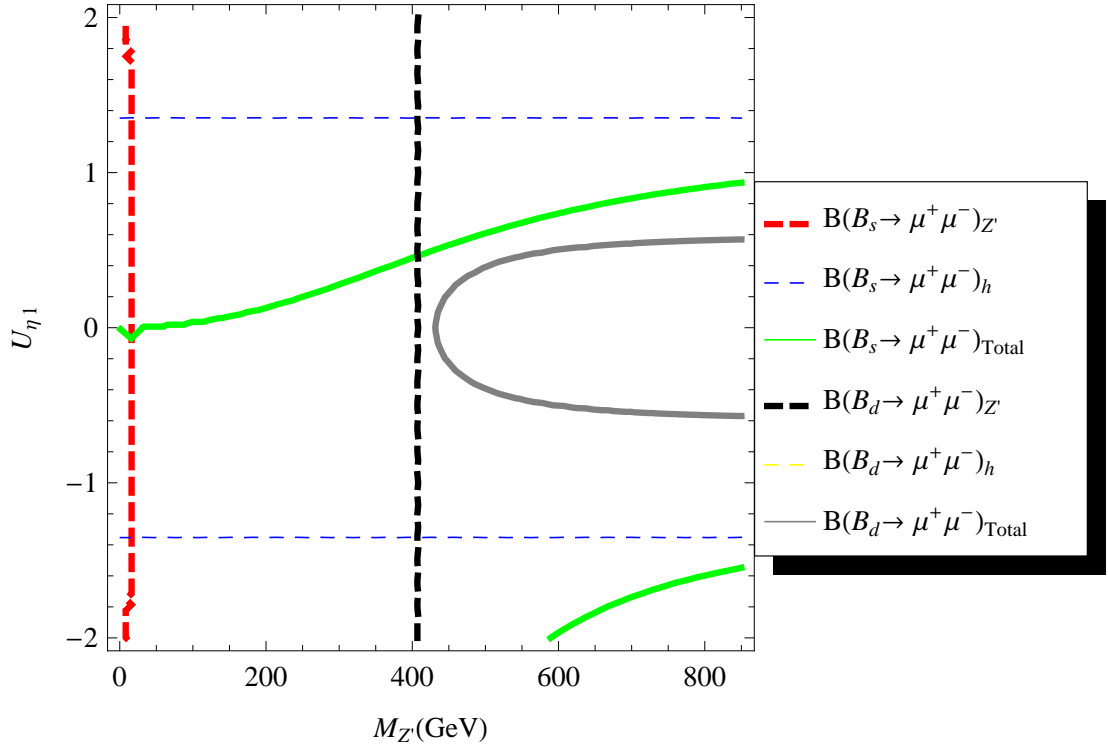}
\caption{Same as Fig.~\ref{fig7} but now for $B_d \rightarrow \mu^+ \mu^-$ and
$B_s \rightarrow \mu^+ \mu^-$ decays.}
\label{fig8}
\end{center}
\end{figure}

\begin{figure}[ht]
\begin{center}
\includegraphics[width=6in]{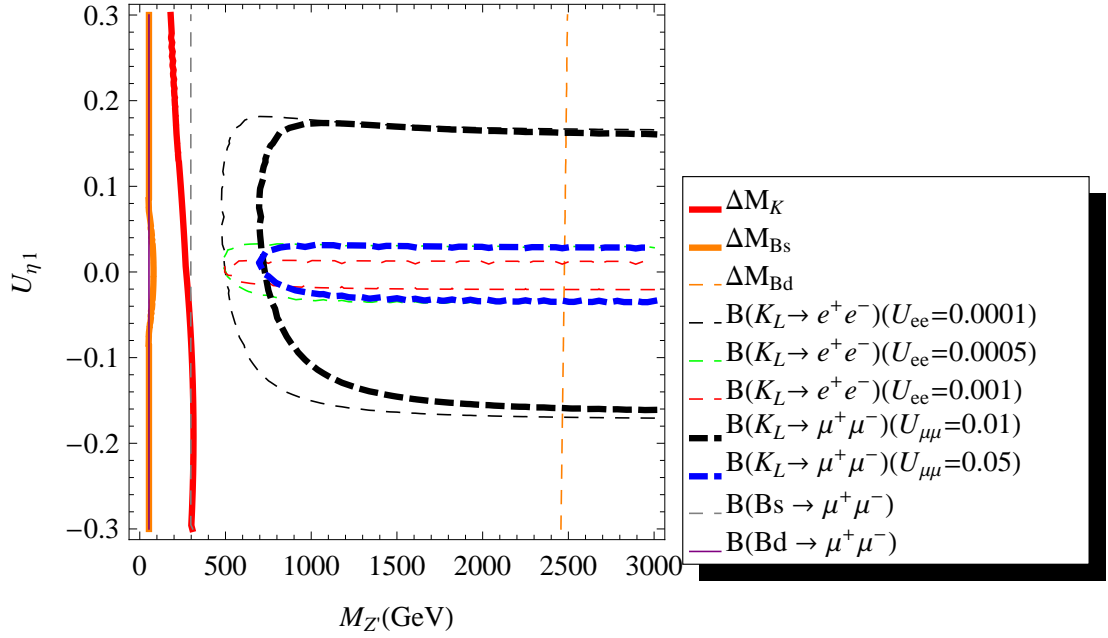}
\caption{This figure summarizes all the previous results
for $K^0-\bar{K}^0$, $B^0-\bar{B}^0$ mass differences and the $K^0$, $B_s$ and $B_d$ decays.}
\label{fig9}
\end{center}
\end{figure}

For the decay $K_L\to\mu\mu$ we use Eq.~(\ref{gamma331}) with $l=l^\prime=\mu$ and $M=K$. In
Fig.~\ref{fig7}, as in Fig.~\ref{fig5}, the red (solid)
vertical line is the contribution of the $Z^\prime$ only and the lower bound on the $Z^\prime$ mass is
around 705 GeV.
As can be seen from Fig.~\ref{fig8}, the scalar contribution does not constrain $U_{\eta1}$. The total
contribution is given by the blue (dashed)
curved line and $M_{Z^\prime}>740$ GeV. Notice from Fig.~\ref{fig7} that this decay has a destructively
interference for $0.01<U_{\eta1}<0.1$ and
constructive for $U_{\eta1}$ outside this region. Finally, see from Fig.~\ref{fig8} that the decay $B_s\to\mu\mu$
does not constrain these parameters anymore.
In Fig.~\ref{fig9} we summarize all constraint when only $Z^\prime$ and $h^0_1$ are considered.

\begin{figure}[ht]
\begin{center}
\includegraphics[width=6in]{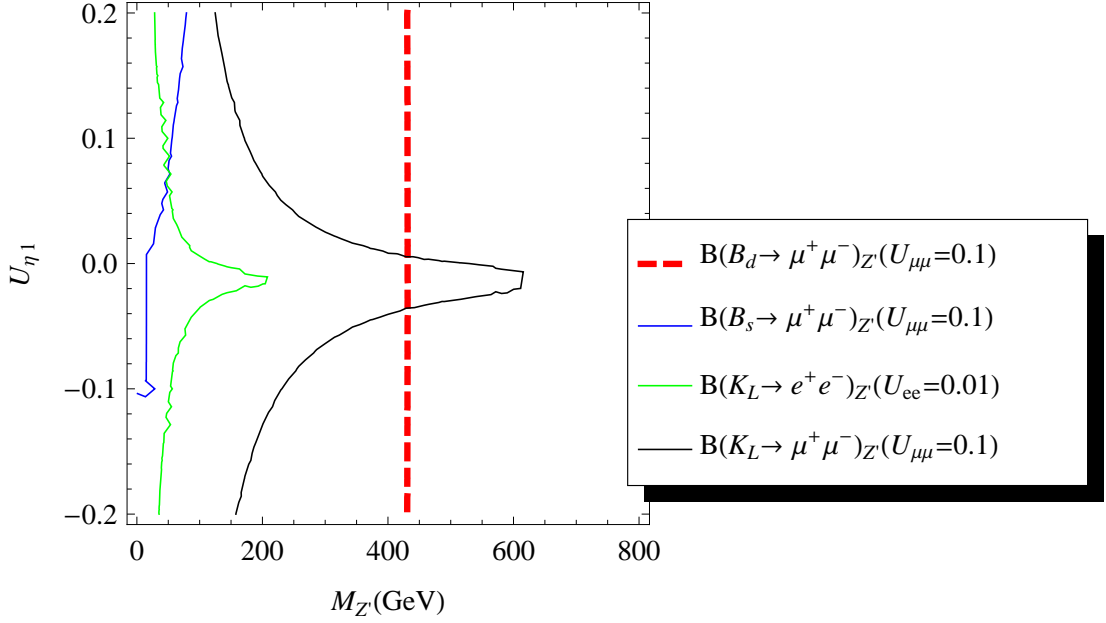}
\caption{Here we consider the contribution of the pseudoscalar to the semi-leptonic decays.  We are assuming that $U_{\eta1}=V_{\eta1}$ and
$U_{\rho1}=V_{\rho1}$ which implies $U_{ee} = U^A_{ee}$ and $U_{\mu \mu} = U^A_{\mu \mu}$ in Eq.~(\ref{gamma331}). The allowed region for the $Z^\prime$ mass and
$U_{\eta 1}$ for fixed values to the pseudo scalar mass at $m_A = 80$ GeV.
The allowed region is always to the right and bounded by the curves and the biggest constraint comes from $K_L \to \mu^+ \mu^-$.}
\label{fig10}
\end{center}
\end{figure}

\begin{figure}[ht]
\begin{center}
\includegraphics[width=6in]{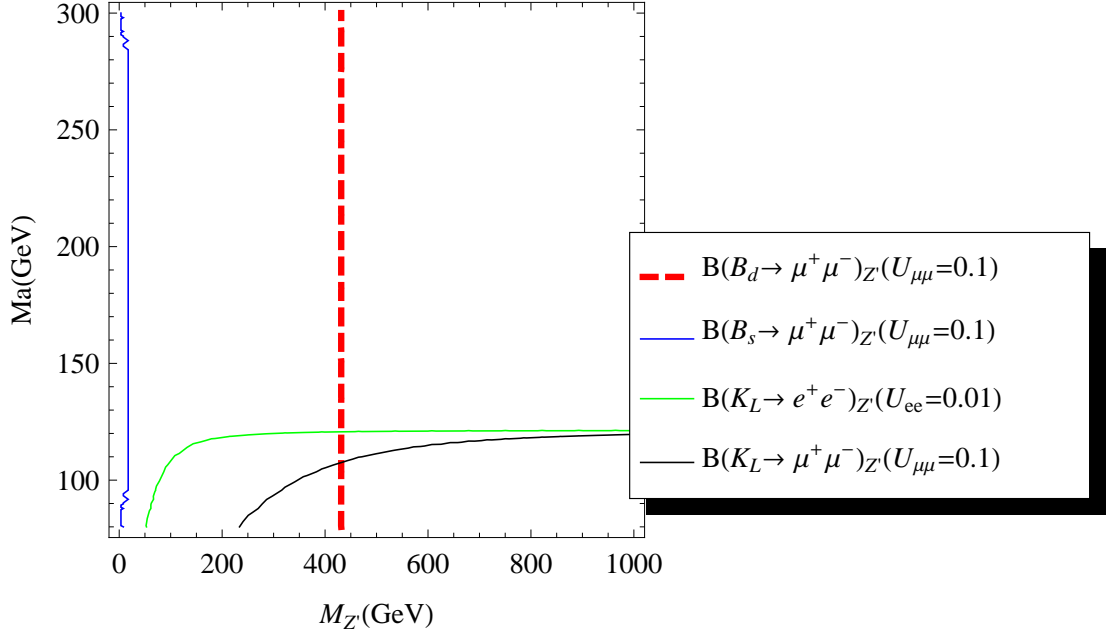}
\caption{Considering the meson semi-leptonic decays including now the contribution of the pseudo scalar,  the allowed region for the $Z^\prime$ and the
pseudoscalar $A^0_1$ masses, for a fixed value of $U_{\eta 1} = V_{\eta1}=0.1$ and $U_{\rho1}=V_{\rho1}=0.42$. The allowed region is always to the
right and bounded by the curves and the biggest constrain came from $K_L \rightarrow \mu^+ \mu^-$.}
\label{fig11}
\end{center}
\end{figure}

The pseudoscalar effects in the leptonic meson decays are  shown in Figs.~\ref{fig10} and \ref{fig11} under the
assumption that $U_{ll^\prime}=U^A_{ll^\prime}$ in Eq.~(\ref{gamma331}), i.e., that $V_{\eta1}=U_{\eta1}$ and  $V_{\rho1}=
U_{\rho1}=0.42$ in Eq.~(\ref{leptons4}).
From the latter equation we see that this is not the more general case but we used it just for the sake of simplicity.

When the nontrivial SM limit discussed in Sec.~\ref{sec:fcnc} is satisfied, $Z$ and $Z^\prime$ decouple i.e., the
respective mixing angle, say $\theta$,
is zero at the tree level. In this case, the masses of the neutral vector bosons are given by
\begin{equation}
M^2_1=\frac{g^2}{4c^2_W}v^2_W\equiv M^2_Z, \quad M^2_2=\frac{g^2}{2c^2_W}\frac{(1-2s^2_W)(4+\bar{v}^2_W)+
s^4_W(4-\bar{v}^4_W)}{1-4s^2_W}v^2_\chi\equiv M^2_{Z^\prime},
\label{masszp}
\end{equation}
where $v^2_W=v^2_\eta+v^2_\rho+2v^2_S$, $\bar{v}_W=v_W/v_\chi$, and $v_S$ is the VEV of the sextet that we
can neglect here. A lower limit of
2.3 TeV for $M_{Z^\prime}$ implies $v_\chi>1.6$ TeV, from Eq.~(\ref{masszp}). On the other hand, since the mass
of the charged vector bosons,
$W^\pm_\mu,V^\pm_\mu, U^{\pm\pm}_\mu$, are given by
\begin{equation}
M^2_W=\frac{g^2}{4}v^2_W,\quad M^2_V=\frac{g^2}{4}(v^2_\eta+2v^2_S+v^2_\chi),\quad M^2_U=
\frac{g^2}{4}(v^2_\rho+2v^2_S+v^2_\chi),
\label{massc}
\end{equation}
with $v_\chi>1.6$ TeV we have $M_V>532$ GeV and $M_U>527$ GeV, using $g^2=4\pi\alpha(Z)$. These values
satisfy the upper bound~\cite{newp}
\begin{equation}
\frac{\sqrt{M^2_V-M^2_U}}{M_W}\leq\sqrt{3}\tan\theta_W,
\label{ub}
\end{equation}
and not $M_V=M_U$, as is the case when we assume $v_\chi\gg v_\eta,v_\rho$ from the very start. Notice that
the exotic charged quarks, which
masses are of the form
$m_j=g_jv_\chi/\sqrt2$, may have masses of the order of 200-300 GeV for reasonable values of the dimensionless
Yukawa couplings $g_j$.

\section{Conclusions}
\label{sec:con}

Here we have considered constraints coming from FCNC processes, $\Delta M_{K,B,D},K_L\to ee,\mu\mu$ and
$B_{d,s}\to \mu\mu$, on the mass of
the $Z^\prime$ neutral vector boson in the m3-3-1 model, taken into account, besides the $Z^\prime$,
the contributions of
the lightest scalar field, $h^0_1$,  which we assume having a mass of 125 GeV, and a pseudoscalar with
arbitrary mass, $m_A$. We first calculated all  entries of the $V^{U,D}_{L,R}$ matrices which modified the Yukawa
couplings in the quarks sector. Next, the matrix elements that relate the symmetry and mass eigenstates in the (pseudo)scalar
sector, ($V_{\eta 1},V_{ \rho1}$) $U_{\eta 1},U_{ \rho1}$,  are fixed by imposing the agreement with the measured mass differences and
branching ratios, on the assumption that $V_{\eta1}=U_{\eta1}$ and $V_{ \rho1}=U_{\rho1}$.
We also have assumed that, the couplings of the scalar $h^0_1$ to the top quark were numerically equal to the coupling
of the Higgs and the top quark in the SM and that the production mechanism was, for  all practical purposes, the same
of the Higgs of the SM as it was discussed in Sec.~\ref{sec:whathiggs}.
In most multi-Higgs models the couplings of $h^0_1$ to other
fermions and to $W$ and $Z$ are not all full strength (i.e., the SM ones) because of the mixing among all the scalar fields
(for an exception see Ref.~\cite{flavour}). In the present model, some of
these couplings may be larger and other smaller than the respective SM values, at least at $\mu=M_Z$.

\begin{figure}[ht]
\begin{center}
\includegraphics[width=5in]{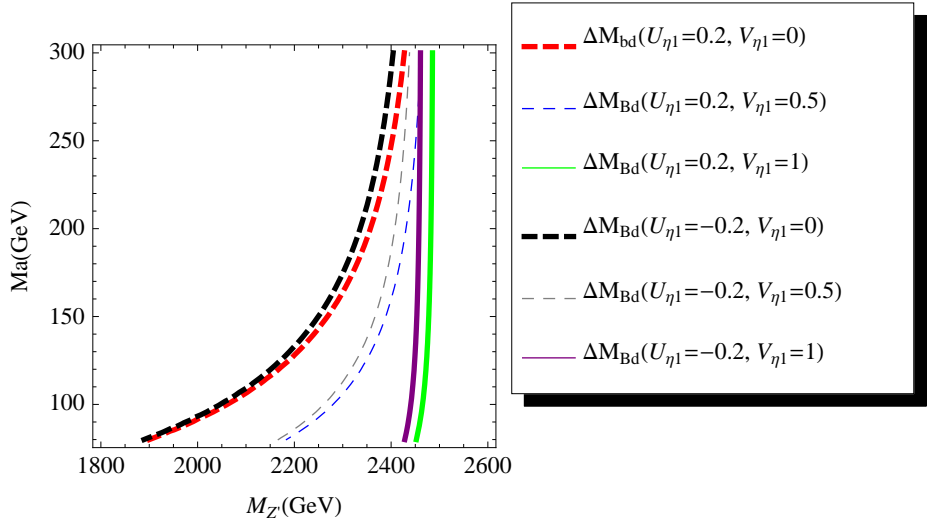}
\caption{Same as Fig.~\ref{fig4} but now considering $V_{\eta1}$ and $V_{\rho1}$ independently of $V_{\eta1}$ and $V_{\rho1}$. Notice that the lower
limit is smaller than in Fig.~\ref{fig4}.}
\label{fig12}
\end{center}
\end{figure}

The amplitude of some of the neutral scalars interfere some times destructively,
as in $\Delta M_{K,D}$, and some times constructively, as  in the $K_L\to ll$ decay. If only $Z^\prime$ is considered,
the lower bound on $M_{Z^\prime}$ from $\Delta M_K$  is $M_{Z^\prime}> 2.5$ TeV and $>27$ TeV in $\Delta M_D$.
When the neutral scalar is considered as well, the constraint is weaker allowing a rather light
$Z^\prime$,  see Secs.~\ref{subsec:mk}, \ref{subsec:mb} and \ref{subsec:mc}.
The strongest constraint on the $Z^\prime$ mass comes from $\Delta B_d$, which is insensible to the scalar contributions
and implies $ m_{Z^\prime}>2.5$ TeV, but when one pseudoscalar is considered it becomes $m_{Z^\prime}>2.3$ TeV if the
the pseudoscalar has a mass of around 180 GeV and under the conditions defined above. However, the latter upper limit
depends on the conditions $V_{\eta1}=U_{\eta1}$ and $V_{ \rho1}=U_{\rho1}$. If this is not the case, i.e., if $V_{\eta1}$ and
$V_{\rho1}$ are considered free parameters, a smaller bound on the $Z^\prime$ mass is obtained:
$M_{Z^\prime}>1.8$~TeV as can be seen in Fig.~\ref{fig12}, which implies $v_\chi>1.27$ TeV, $M_V>412$ GeV, and $M_U>419$ GeV.
The leptonic kaon decay into two leptons implies a lower bound for this mass of 740 GeV, see Sec~\ref{sec:deltaf1} for discussions.

A final remark is in order here. From Eqs.~(\ref{yukan}),  (\ref{kapasd}) and (\ref{kapasu}), we see that the constraints depend on the
matrix elements of $V^q_{L,R}$ given in Eqs.~(\ref{vudl331}) and (\ref{vudr331}).
These matrices have to diagonalize the quark mass matrices and hence
they depend on the input parameters
$G,G,\tilde{G},\tilde{F}$ and VEVs, in these mass matrices
In this work we found a set of parameters that is compatible with the
quark masses, at $\mu=M_Z$, and the CKM matrix. There could exist a different set,
i.e., a different quark mass matrix, showing the same compatibility,
which will be diagonalized by different $V^q_{L,R}$ matrices and,
therefore, resulting in different values for the $Z^\prime$ mass
constraint. The set we found is show below Eq.(\ref{vudr331}). We tried to find a
different one without success. It seems that finding another set is not
a trivial task, but it can, in principle, exist.
There may be solutions with a heavy $Z^\prime$ when there is no destructive interference in the $\Delta M_K$ amplitude but there is in $\Delta M_{B_d}$, and so on.
The main result of our work is shown that the interference between $Z^\prime$ and (pseudo)scalar fields exist in some range of the parameters.
Hence, the effects considered here may be at work
in $Z^\prime$-searches at the LHC as well but the interference will be with heavy (pseudo)scalars, different from $h^0_1$.

It is well known that the magnetic dipole transitions $b\to s+\gamma$,
or $b\to d+\gamma$, have branching ratios of the order of $10^{-4}$ and are in agreement with the SM
predictions~\cite{babar}. For a recent analysis see Ref.~\cite{promberger2}. In the present model this sort of decays and CP violation
also arise at  the one-loop order through penguin and box diagrams. However, in the present case there are contributions of
the singly and doubly charged scalars, exotic quarks, and singly and doubly charged vector bosons present in
the model. The same happen with the $\Delta F=2$ processes since there are box diagrams involving singly and doubly
charged scalar and vector bosons and exotic quarks as well. These contributions to $\vert\Delta F\vert=1,2$ will be considered elsewhere.

The search for a $Z^\prime$-like resonance has been done at the LHC. However, as in previous searches, the
results are usually obtained in the context of a given model. For instance, in a top color assisted spontaneous symmetry breaking
scenario this sort of (leptophobic) resonance has been excluded for $M_{Z^\prime}<1.3$ TeV, if $\Gamma_{Z^\prime}=
0.012M_{Z^\prime}$, and $M_{Z^\prime}<1.9$ TeV, if $\Gamma_{Z^\prime}=0.10M_{Z^\prime}$~\cite{zprimecms}. Notwithstanding,
the application of these bounds to the model  considered here is not straightforward and has to be done in a separate work.

Last but not least, we would like to say that the m3-3-1 solution that we have presented here can be falsifiable
in the near future: When the strength of the $VVh^0_1,\;V=W,Z$  where measured, given at least upper limits for $U_{\eta1}$ and
$U_{\rho1}$, then we can check if all the couplings of the 125 GeV Higgs boson with the gauge bosons and all the fermions, when measured
with sufficient precision, agree or not with those in Eqs.~(\ref{kapasd}) and (\ref{kapasu}) when $\eta^0$ and $\rho^0$ are projected on
$h^0_1$.

\acknowledgments

The authors would like to thank CAPES for full support
(A. C. B. M.) and CNPq and FAPESP for partial support (V.P.)



\newpage

\end{document}